\documentclass[twocolumn]{aastex631}
\usepackage{amssymb,amsmath,amsfonts}
\usepackage{graphicx}
\usepackage{xcolor}
\usepackage{verbatim}
\usepackage{color}
\usepackage{mdframed}
\usepackage{soul} 
\usepackage{braket}
\usepackage{enumitem}
\usepackage[percent]{overpic}
\usepackage{float}
\let\tablewidth\relax  
\usepackage{tabularray}
\usepackage{booktabs}
\renewcommand{\arraystretch}{1.6}
\usepackage{soul}
\include{eps2pdf}
\usepackage{fontawesome}
\usepackage{sidecap}
\usepackage{multirow}
\usepackage[splitrule,bottom]{footmisc} 
\allowdisplaybreaks

\def\d{\partial}

\newcommand{\av}[1]{\left\langle{#1}\right\rangle} 
\newcounter{sarrow}

\newcommand{\test}[1]{{\sc \mathtt{TEST#1} }}
\newcommand{\simbig}{{\sc SimBIG }}

\def\gpch{{h^{-1}\rm{Gpc}}}
\def\mpch{{h^{-1}\rm{Mpc}}}
\def\hmpc{{{\rm Mpc}^{-1} h}}

\def\bx{{\bf x}}

\def\bk{{\bf k}}
\def\bq{{\bf q}}

\def\hk{{\hat k}}

\def\cD{{\mathcal D}}

\def\cP{{\mathcal P}}
\def\cS{{\mathcal S}}

\definecolor{myblue}{rgb}{0.1, 0.4, 0.5}
\definecolor{color1}{rgb}{0.5, 0.5, 0.4}
\definecolor{lightpurple}{HTML}{9999FF}
\definecolor{lightblue}{HTML}{66CCFF}
\definecolor{sweetblue}{HTML}{33CCFF}
\definecolor{waterblue}{HTML}{0099CC}
\definecolor{steelblue}{HTML}{6699CC}
\definecolor{kingsblue}{HTML}{1CBCAC}
\definecolor{forest}{HTML}{006175}
\definecolor{oldgreen}{HTML}{008E3E}
\definecolor{saddlebrown}{HTML}{BC4C1C}
\definecolor{ube}{HTML}{301728}

\shorttitle{{\sc SimBig}: Skew Spectra of BOSS-CMASS Galaxies}
\shortauthors{Hou \&  Moradinezhad Dizgah et al.}

\begin{document}

\title{\simbig: Cosmological Constraints from the Redshift-Space Galaxy Skew Spectra}

\author{Jiamin Hou}
\altaffiliation{jiamin.hou@ufl.edu}
\affiliation{Department of Astronomy, University of Florida, 211 Bryant Space Science Center, Gainesville, FL 32611, USA}
\affiliation{Max-Planck-Institut f\"ur Extraterrestrische Physik, Postfach 1312, Giessenbachstrasse 1, 85748 Garching bei M\"unchen, Germany}

\author{Azadeh Moradinezhad Dizgah}
\altaffiliation{azadeh.moradinezhad@lapth.cnrs.fr}
\affiliation{Laboratoire d’Annecy de Physique Théorique (CNRS/USMB), F-74940 Annecy, France}
\affiliation{Département de Physique Théorique, Université de Genève, 24 quai Ernest Ansermet, 1211 Genève 4, Switzerland}

\author{ChangHoon Hahn}
\affiliation{Department of Astrophysical Sciences, Princeton University, Princeton NJ 08544, USA} 


\author{Michael Eickenberg}
\affiliation{Center for Computational Mathematics, Flatiron Institute, 162 5th Avenue, New York, NY 10010, USA}

\author{Shirley Ho}
\affiliation{Center for Computational Astrophysics, Flatiron Institute, 162 5th Avenue, New York, NY 10010, USA}

\author{Pablo Lemos}
\affiliation{Department of Physics, Universit\'{e} de Montr\'{e}al, Montr\'{e}al, 1375 Avenue Th\'{e}r\`{e}se-Lavoie-Roux, QC H2V 0B3, Canada}
\affiliation{Mila - Quebec Artificial Intelligence Institute, Montr\'{e}al, 6666 Rue Saint-Urbain, QC H2S 3H1, Canada}
\affiliation{Center for Computational Mathematics, Flatiron Institute, 162 5th Avenue, New York, NY 10010, USA}

\author{Elena Massara}
\affiliation{Waterloo Centre for Astrophysics, University of Waterloo, 200 University Ave W, Waterloo, ON N2L 3G1, Canada}
\affiliation{Department of Physics and Astronomy, University of Waterloo, 200 University Ave W, Waterloo, ON N2L 3G1, Canada}

\author{Chirag Modi}
\affiliation{Center for Computational Mathematics, Flatiron Institute, 162 5th Avenue, New York, NY 10010, USA}
\affiliation{Center for Computational Astrophysics, Flatiron Institute, 162 5th Avenue, New York, NY 10010, USA}

\author{Liam Parker}
\affiliation{Center for Computational Astrophysics, Flatiron Institute, 162 5th Avenue, New York, NY 10010, USA} 

\author{Bruno R\'egaldo-Saint Blancard}
\affiliation{Center for Computational Mathematics, Flatiron Institute, 162 5th Avenue, New York, NY 10010, USA}

\begin{abstract}
Extracting the non-Gaussian information of the cosmic large-scale structure (LSS) is vital in unlocking the full potential of the rich datasets from the upcoming stage-IV galaxy surveys. Galaxy skew spectra serve as efficient beyond-two-point statistics, encapsulating essential bispectrum information with computational efficiency akin to power spectrum analysis. This paper presents the first cosmological constraints from analyzing the full set of redshift-space galaxy skew spectra of the data from the SDSS-III BOSS, accessing cosmological information down to nonlinear scales. Employing the \simbig forward modeling framework and simulation-based inference via normalizing flows, we analyze the CMASS-SGC sub-sample, which constitute approximately 10\% of the full BOSS data. Analyzing the scales up to $k_{\rm max}=0.5 \ \mpch$, we find that the skew spectra improve the constraints on $\Omega_{\rm m}, \Omega_{\rm b}, h$, and $n_s$ by 34\%, 35\%, 18\%, 10\%, respectively, compared to constraints from previous \simbig power spectrum multipoles analysis, yielding $\Omega_{\rm m}=0.288^{+0.024}_{-0.034}$, $\Omega_{\rm b}= 0.043^{+0.005}_{-0.007}$, $h=0.759^{+0.104}_{-0.050}$, $n_{\rm s} = 0.918^{+0.041}_{-0.090}$ (at 68\% confidence limit). On the other hand, the constraints on $\sigma_8$ are weaker than from the power spectrum. Including the Big Bang Nucleosynthesis (BBN) prior on baryon density reduces the uncertainty on the Hubble parameter further, achieving $h=0.750^{+0.034}_{-0.032}$, which is a 38\% improvement over the constraint from the power spectrum with the same prior. Compared to the \simbig bispectrum (monopole) analysis, skew spectra offer comparable constraints on larger scales ($k_{\rm max}<0.3\ \mpch$) for most parameters except for $\sigma_8$.
\end{abstract}


\section{Introduction}

The LSS holds valuable clues about the Universe's origin, composition, and evolution, making it a crucial arena for testing fundamental physics. In the coming years, the next-generation galaxy surveys including Dark Energy Spectroscopic Survey~\citep[DESI;][]{DESI:2016fyo}, {\it Euclid}~\citep[][]{EUCLID:2011zbd}, Nancy Roman Space Telescope~\citep[Roman;][]{Wang2022:RomanHLSS}, SPHEREx~\citep[][]{SPHEREx2018}, and Subaru Prime Focus Spectrograph~\citep[PFS;][]{Tamura:2016wsg} will provide a detailed spectroscopic view of the three-dimensional distribution of galaxies across significantly larger cosmic volumes than current surveys. This expanded scope promises a marked reduction in statistical uncertainties in the measured clustering statistics of galaxies, yielding unparalleled constraints on the standard $\Lambda$CDM model and opening up exciting new avenues for exploring new physics. 

The nonlinear nature of the LSS implies that the commonly applied two-point correlation functions are insufficient in capturing the full cosmological information of galaxy clustering. Therefore, to exploit the upcoming galaxy surveys to their full potential, constructing summary statistics that capture the non-Gaussian clustering information is crucial. This has spurred a significant amount of work in the literature focusing on constructing optimal summary statistics, including bispectrum \citep{Scoccimarro2015:bkRSD,Sugiyama2019:BkRSD,DAmico2022:BOSSbispec,Philcox:2021kcw}, skew spectra~\citep{Schmittfull:2014tca,MoradinezhadDizgah:2019xun,Schmittfull2020:SSrsd,Hou:2022rcd}, marked 2-point functions~\citep{Sheth2005:mark,White2016:mark,Massara2021:markedPk}, density split statistics~\citep{Paillas2021:DensitySplit}, and wavelet scattering transforms~\citep{Eickenberg:2022qvy,Valogiannis2022:WST}. More ambitious approaches to extract information at the field level without relying on any summary statistics~\citep{Jasche2019:FM,Lavaux2019:FMBOSS,Andrews:2022nvv,Kostic2023:FLIeft,Stadler2023:FLIRSD} have also been a topic of growing interest. The potential of these approaches has been extensively explored using Fisher forecasts ({\it e.g.}, \citealt{Agarwal:2020lov,Hahn:2020lou, Valogiannis:2021chp,Eickenberg:2022qvy,Massara:2022zrf,Hou:2022rcd,Paillas2023:DSfisher}), and very recently, several of them ({\it e.g.}, ~\citealt{Paillas2023:DSboss,Hahn2023:simbigBk,Blancard2023:simbigWST,Lemos2023:simbigCNN,Valogiannis:2023mxf}) have been applied to the existing data from Sloan Digital Sky Survey (SDSS)-III Baryon Oscillation Spectroscopic Survey Data Release 12~\citep[BOSS DR12;][]{Reid2016:BOSSDR12,BOSSCollaboration2017}. 

Among various proposed summary statistics, the galaxy skew spectra \citep{Schmittfull:2014tca,MoradinezhadDizgah:2019xun,Dai:2020adm,Schmittfull2020:SSrsd,Chakraborty:2022aok} are optimal, yet computationally efficient proxy statistics for the galaxy bispectrum. They are constructed from the cross-correlations of the observed galaxy density field with appropriately weighted quadratic galaxy fields. For parameters such as galaxy biases, the growth rate of structure, and the amplitude of primordial non-Gaussianity (PNG), by construction, the skew spectra fully capture the information of the bispectrum in the linear regime \citep{MoradinezhadDizgah:2019xun}. Despite these encouraging results from Fisher forecasts \citep{MoradinezhadDizgah:2019xun, Hou:2022rcd}, the applicability of the full redshift space galaxy skew spectra to observational data, {in particular down to nonlinear scales}, has not been studied. This investigation is the focus of the current article.

The standard approach to cosmological inference from the LSS is based on a Bayesian likelihood analysis and requires: (1) a theoretical model of the observable, (2) an explicit form of the likelihood, and (3) schemes for correcting the observational systematics in the measurements. These requirements can hinder the maximal exploitation of upcoming LSS data. 

First, if using perturbation theory (PT) to describe the observables, the limited validity of the perturbative description restricts the theory-based analyses to relatively large scales ($k_{\rm max}\simeq 0.25 \, \hmpc$).\footnote{For analysis in which simulation-based emulators are used to go beyond the perturbative limit ({\it e.g.}, \cite{Yuan:2022jqf,Paillas:2023cpk,Valogiannis:2023mxf}), the emulator error is the limiting factor. The increasing precision of the data imposes more stringent constraint on the required accuracy of both PT- and emulator-based modeling.} Despite the considerable recent developments in perturbative description of the LSS \citep{McDonald:2006mx, Baumann:2010tm, Carrasco:2012cv, Senatore:2014eva,Senatore:2014via,Senatore:2014vja, Vlah:2015sea, Blas:2015qsi, Ivanov:2018gjr, Cabass:2022avo}, the PT-based analyses are intrinsically limited to the weakly nonlinear regime. Moreover, apart from the galaxy bispectrum, for some of the powerful recently proposed statistics such as wavelet scattering transforms, a perturbative description is missing altogether. 

The second limitation of the standard analysis is that the assumption of a Gaussian likelihood, which relies on the central limit theorem, does not hold on large scales with low signal-to-noise ratio and on small scales where modes are highly correlated. Thus an explicit assumption of the Gaussian likelihood, can potentially bias the parameter constraints \citep{Sellentin:2017fbg,Hahn:2018zja}.

Lastly, the standard analysis accounts for observational systematics such as targeting, imaging, and spectroscopic incompleteness by applying weights designed to correct their effects on measured observables~\citep{ross2012, ross2017, kitanidis2020}. However, even at the precision level of current surveys, the weights are insufficient at correcting for effects such as fiber collisions, which can bias the clustering measurements on weakly non-linear scales~\citep{guo2012, hahn2017a, bianchi2018}.
Moreover, it is important to note that these correction techniques are currently optimized only for two-point statistics and their validity for other summary statistics has not been thoroughly tested.

To overcome these challenges, simulation-based inference~\citep[SBI;][]{Cranmer_2020,Lueckmann:sbi} is emerging as a promising avenue.\footnote{The SBI is also commonly referred to in the literature as {\it likelihood-free} or {\it implicit likelihood} inference.} Without a strong assumption for the form of the likelihood, SBI enables estimation of the parameter posteriors (or the likelihood function) by leveraging high-fidelity numerical simulations to forward model the observables and employing deep generative models from machine learning for efficient parameter inference. In the context of galaxy clustering, several recent studies have explored the potential of SBI, employing either a set of summary statistics or at the field-level ({\it e.g.}, \citealt{deSanti:2023zzn,deSanti:2023rsw,Tucci:2023bag,Modi:2023llw}). In terms of application of the SBI to observational data, \cite{Hahn2022:simbigPk, Hahn2022:simbigMockChallenge}
introduced Simulation-Based Inference of Galaxies (\textsc{SimBIG}), a forward-modeling framework that enables performing SBI on a spectroscopic galaxy sample, employing either summary statistics or at the field-level. The \simbig forward model capitalizes on high-fidelity $N$-body simulations, coupled with the state-of-the-art Halo Occupation Distribution (HOD) model to accurately simulate nonlinear galaxy clustering. Notably, \simbig integrates pertinent observational effects such as survey masks and fiber collisions. In contrast to the common approach of accounting for the fiber collisions via correction weights, \simbig forward models their effect. In a series of recent papers, \simbig has been used to obtain cosmological constraints from SDSS-III BOSS CMASS galaxies. By employing several statistics (including the power spectrum \cite{Hahn2022:simbigPk,Hahn2022:simbigMockChallenge}, the bispectrum \cite{Hahn2023:simbigBk}, and the wavelet scattering transforms \cite{Blancard2023:simbigWST}), and by performing field-level analysis using convolutions neural networks \citep{Lemos2023:simbigCNN}, these works have highlighted the substantial potential of SBI in obtaining constraints that surpass the standard analysis of galaxy clustering~\citep{Hahn2023:simbigWave2}.

This is the first paper using the \simbig framework to obtain cosmological constraints from analyzing the full set of redshift-space galaxy skew spectra, accessing information down to nonlinear scales. As part of this analysis, we also present a new estimator for the skew spectra that extends the previously developed estimators for periodic-box simulations \citep{Schmittfull2020:SSrsd,Hou:2022rcd} to account for survey mask and angular systematic weights so that it can be used on observations.

The structure of the paper is as follows: \S\ref{sec:skewspec} provides an overview of the galaxy skew spectra and the new estimator for measuring the skew spectra on a realistic galaxy sample. In \S \ref{sec:boss}, we briefly describe the observational data used in our analysis, and in \S\ref{sec:simbig} we describe the \simbig pipeline (the forward modelling, the simulation-based inference, and the validation tests). We show the results in \S\ref{sec:results}, including the validation tests on synthetic data and the inferred posterior distribution of cosmological parameters from the BOSS-CMASS sample. Finally, we draw our conclusions in \S\ref{sec:conclude}. 

We provide additional details and complementary results in a series of appendices. In Appendix \ref{app:kernels}, we provide explicit forms of the skew spectra, in Appendix \ref{app:full}, we discuss the constraints on HOD parameters, and in Appendix \ref{app:further_studies} we investigate the dependence of the inferred constraints on several analysis choices including the outlier removal scheme, the removal of shot noise, the scale cuts, and the smoothing scale. 

\section{Summary Statistics:\\ Galaxy Skew Spectra}
\label{sec:skewspec}

The galaxy skew spectra are efficient proxy statistics for the bispectrum and offer several advantages over the bispectrum; first, they are simple to interpret since they are functions of a single wavenumber and not triangle shapes. Second, the computational cost of measuring them from the data is comparable to that of the power spectrum. While accounting for all bispectrum triangles requires $\mathcal O(N^2)$ operations capturing the full information of the bispectrum using the skew spectra requires ${\mathcal O}(N \log N)$ operations, where $N = (k_{\rm max}/\Delta k)^3$ is the number of 3D Fourier-space grid points at which the fields are evaluated. Third, for standard Bayesian inference, if estimating the covariance matrices from mocks, the skew spectra requires a significantly smaller number of mocks given the significantly smaller size of the data vector. Lastly, in comparison to the bispectrum \cite{Pardede:2022udo}, accounting for the survey window function is expected to be considerably simpler for the skew spectra~\citep{Hou:2022aaa}.

\subsection{Theoretical Construction}
The skew spectra are constructed by cross-correlating the observed galaxy density field, $\delta_g$, with an appropriately weighted square of it, $\cS_n$~\citep{Schmittfull:2014tca,Schmittfull2020:SSrsd};
\begin{equation}\label{eqn:sk_def}
    \cP_{\cS_n \delta}(k) 
    = \int \frac{d\hk}{4\pi} \av{\delta_{\rm g}(-\bk) \cS_n(\bk)}.
\end{equation}
The quadratic field, $\cS_n$, is a product of two fields in configuration space (evaluated at the same spacial positions), thus, a convolution in Fourier space;
\begin{equation}
    \cS_n(\bk) = \int_{\bf q} \cD_n(\bq, \bk-\bq) \delta^R_{\rm g}(\bq) \delta^R_{\rm g}(\bk-\bq),
\end{equation}
with $\int_{\bf q}\equiv \int d^3 q / (2\pi)^3$.
The fields involved in the convolution are smoothed, $\delta_g^R(\bk) = \delta_g(\bk) W_R(\bk$), to limit the contribution of small-scale fluctuations to skew spectra. While a top-hat filter in Fourier space amounts to imposing a sharp $k$-cut on contribution of small-scale modes to the convolution integral, to avoid edge effects in the measurements of the skew spectra, we use a Gaussian smoothing kernel $W_R(\bk) = \exp(-k^2 R^2/2)$. In perturbative analysis of the skew spectra \citep{Schmittfull:2014tca, MoradinezhadDizgah:2019xun, Schmittfull2020:SSrsd}, smoothing the field (with relatively large $R$) ensures that only the Fourier modes on the semi-linear regime are included. For SBI performed in this paper, the smoothing serves the purpose of limiting the analysis to scales where the forward model is most trusted.\footnote{Similarly, in numerical Fisher forecasts \citep{Hou:2022rcd} the smoothing ensures that only modes on scales not affected by the resolution of the simulations are included in the analysis.} 

The weights, $\cD_n(\bq, \bk-\bq)$, are constructed such that for a theoretical bispectrum that can be decomposed as a sum of product-separable contributions (in terms of the three Fourier modes forming a triangle), the skew spectra correspond to the maximum likelihood estimators for parameters appearing as overall amplitudes of individual contributions in the sum. Explicit forms of the kernels $\cD_n(\bq, \bk-\bq)$ are given in Appendix \ref{app:kernels}. This implies that, by construction, considering full set of 14 galaxy skew spectra in redshift space (3 in real space) fully capture the information of the bispectrum on large scales for parameters such as galaxy biases, the amplitude of primordial power spectrum and bispectrum, and the growth rate of structure \citep{Schmittfull:2014tca, MoradinezhadDizgah:2019xun}.\footnote{We define the {\it full set} of redshift-space galaxy skew spectra by including the 14 kernels that encode the line-of-sight velocity information, in contrast to a {\it subset} of skew spectra that are constructed from the 3 real-space kernels.} For other cosmological parameters (including $\Lambda$CDM parameters and the sum of neutrino masses), by performing numerical Fisher forecasts on \textsc{Quijote} simulations \citep{Villaescusa-Navarro:2019bje}, \cite{Hou:2022rcd} demonstrated that the skew spectra can considerably improve the constraints from the power spectrum, at the level comparable to the bispectrum.

In addition to the clustering component in Eq. \eqref{eqn:sk_def}, each of the skew spectra receives a shot noise contribution, which in the Poisson limit is given by
\begin{eqnarray}\label{eqn:skewspec_shot} \vspace{-0.2in}
    \mathcal{P}_{\cS_n \delta}^{\text {shot}}(k)
    \!=\!
    \frac{1}{2}\! \int\!\! d \mu_k \!\Bigg\{\left[\frac{1}{\bar{n}^2}+\frac{P_g(\bk)}{\bar{n}}\right] \!J_{\cD_n}(\bk)
    \!+\!\frac{2}{\bar{n}} \tilde{J}_{\cD_n}(\bk)\Bigg\},\nonumber\\,
\end{eqnarray}
where $P_g(\bk)$ is the full 3D galaxy power spectrum and the kernels $J_{\cD_n}$ and $\tilde{J}_{\cD_n}$ are given by 

\begin{eqnarray} 
\hspace{-0.1in}&& J_{\cD_n}(\bk)\!=\! \int_{\bf q}  W_R(\mathbf{q}) W_R(\bk-\mathbf{q}) \cD_n(\mathbf{q}, \bk-\mathbf{q}), \\
\hspace{-0.1in}&& \tilde{J}_{\cD_n}(\bk)\!=\! \int_{\bf q}\! W_R(\mathbf{q}) W_R(\bk-\mathbf{q}) D_n(\mathbf{q}, \bk-\mathbf{q}) P^R_{\rm g}(\bq).
\end{eqnarray}


\subsection{Estimators for Survey Data}
Previous works have focused on the analysis of the skew spectra on simulated dark matter distribution and halo catalogs in $N$-body simulations with periodic boundary conditions~\citep{Schmittfull2020:SSrsd,Hou:2022rcd}. In order to analyze the data from BOSS survey, we extend the previous estimators to measure the skew spectra using on Fast Fourier Transforms (FFT) \citep{Schmittfull2020:SSrsd,Hou:2022rcd} to incorporate the survey geometry and observational systematics. We thus construct an equivalent of Feldman-Kaiser-Peacock \citep[FKP;][]{Feldman:1993ky} estimators for the power spectrum and bispectrum, where the galaxy sample is characterized by $n_g$ galaxies at positions $\bx$ and a randoms catalog with $n_r$ objects to incorporate the radial and angular selection functions, 
\begin{equation}\label{eqn:fkp}
    \delta_{\rm g}(\bx) = {I_{33}}^{-1/3} w_{\rm fkp}(\bx) \big[n'_{\rm g}(\bx) - \alpha\, n'_{\rm r}(\bx)\big].
\end{equation}
Here, $n_{\rm g}'(\bx)=w_{\rm c}(\bx)n_{\rm g}(\bx)$ and $n_{\rm r}'(\bx)=w_{\rm c,r}(\bx) n_{\rm r}(\bx)$ are the weighted number densities of the observed and random galaxies, and $\alpha \ (\ll 1)$ is their ratio. As in previous \simbig \ analyses, for each observed galaxy, we assign the completeness weight of $w_{\rm c}= w_{\rm sys} w_{\rm noz}$, where $w_{\rm sys}$ is an angular systematic weight based on stellar density and seeing conditions, and $w_{\rm noz}$ is a redshift failure weight. Since for the fiber collisions, the standard weighting scheme has been shown to be inaccurate \citep{Hahn:2016kiy}, we forward model them in our pipeline instead of using weights as is commonly done in other analyses of BOSS data (for further details, see~\citealt{Hahn2022:simbigMockChallenge}).  We have also included the FKP weight $w_{\rm fkp}(\bx) = \left[1+\bar n(g)P_0\right]^{-1}$ (assuming $P_0 = 10^4 \,[\mpch]^3$) to reduce the variance of the estimator~\citep{Feldman1994:fkp}.\footnote{We adopted the FKP weights minimizing the variance for the power spectrum. This may not be optimal for skew spectra. We leave the extension for future work.} The normalization factor of $I_{33}$ in Eq. \eqref{eqn:fkp} is defined in analogy to that of the bispectrum~\citep{Scoccimarro2015:bkRSD} since both skew spectra and bispectrum have three powers of galaxy density fields,
\begin{eqnarray}\label{eqn:delta_g}
    I_{33} &=& \int d^3 \bx\, \bar{n}_{\rm g}^3(\bx) w^3_{\mathrm{c}}(\bx) w^3_{\mathrm{fkp}}(\bx) \nonumber \\
    &\equiv& \sum_{i=1}^{N_{\rm g}} \bar{n}_{{\rm g}}^2(\bx_i) w^2_{\mathrm{c}}(\bx_i) w^3_{\mathrm{fkp}}(\bx_i),
\end{eqnarray}
where $\bar{n}_{\rm g}$ is the mean number density per redshift bin for the observed galaxy galaxies. 

As a simplifying assumption, we consider a fixed line-of-sight (LoS) direction when constructing the quadratic fields. While in computing the cross-correlation between the linear and the quadratic field the varying LoS is accounted for as in the~\cite{Yamamoto2006:estimator} estimator. The skew spectra correspond to the monopole of the cross-correlation between the input galaxy field and its square. 

For implementation, we used the {\sc nbodykit}~\citep{Hand:2017pqn} \href{https://github.com/bccp/nbodykit}{\faGithub}\footnote{\url{https://github.com/bccp/nbodykit}}. 
We apply the weights $\cD_n(\bq, \bk-\bq)$ (see Eq.~\ref{eqn:sk_def}) to the first two linear density fields to obtain the quadratic field. We then prepare both the quadratic field and the third linear density field as {\sc FKPCatalog} objects, an {\sc nbodykit}-based interface for simultaneous modeling of a data and a random catalog. These two {\sc FKPCatalog} objects are then passed to the {\sc ConvolvedFFTPower} module for cross-correlating the quadratic field and the linear field.

In analogy to the power spectrum and the bispectrum FKP estimators, we subtract the Poisson shot noise from the measured skew spectra. The power-spectrum-independent contribution to shot noise in Eq. \eqref{eqn:skewspec_shot} is estimated by inserting two unity fields, whereas the kernels that depend on the power spectrum are computed by correlating a smoothed 3D power spectrum $P^R_{\rm g}(\bq)=W_R(\bq) P_g(\bq) $ and a unity field. Due to the survey geometry, the full 3D power spectrum is non-Hermitian, whereas the unity fields are real-valued Hermitian fields and the third dimension of the field is halved after the FFT. In practice, we concatenate two hermitian unity fields to match the dimensionality. 

As in the previous \simbig analysis of the galaxy power spectrum and bispectrum, in addition to the skew spectra, we also include the average galaxy number density, $\bar n_g$. Therefore, in total, the data vector that we pass to the SBI pipeline has 1401 elements, with $k_{\rm min}=0.003 \hmpc$ that corresponds to the fundamental frequency of the BOSS SGC, and $\Delta k = 0.005 \,\hmpc$.\footnote{We note that the choice of the fundamental frequency could be susceptible to systematic effects approaching the survey boundary as we discuss in \S Appendix \ref{app:further_studies}} We explore the impact of using a different $k_{\rm min}$ and its implications on our analysis.

\section{Observational Data: \\ BOSS CMASS Galaxy Sample}
\label{sec:boss}

As in previous \simbig analyses \citep{Hahn2022:simbigPk, Hahn2022:simbigMockChallenge, Hahn2023:simbigBk, Blancard2023:simbigWST, Lemos2023:simbigCNN, Hahn2023:simbigWave2}, we consider the CMASS galaxy sample (consisting mainly of luminous red galaxies) of BOSS DR12 \citep{SDSS:2011jap,BOSS:2012dmf}. Since the \simbig \ forward model employs \textsc{Quijote} $N$-body simulations with the box size of $L_{\rm box} = 1 \ {\rm Gpc}^{-1}h$ to model the nonlinear clustering of dark matter, we only use the data in the southern galactic cap (SGC) and further apply angular cuts of ${\rm DEC} > -6^\circ$ and $-25^\circ<{\rm RA}<28^\circ$ and a radial cut corresponding to redshift range of $0.45<z<0.6$. 
The resulting galaxy sample spans 
$\sim 10\%$ of the full BOSS data and about 50\% of the CMASS-SGC sample,
 with 109,636 galaxies and a number density of $\sim\!4.2\times 10^{-4}\, [{\rm Mpc}^{-1}h]^3$. 


\begin{figure*}[htbp!]
    \centering
    \includegraphics[width=\textwidth]{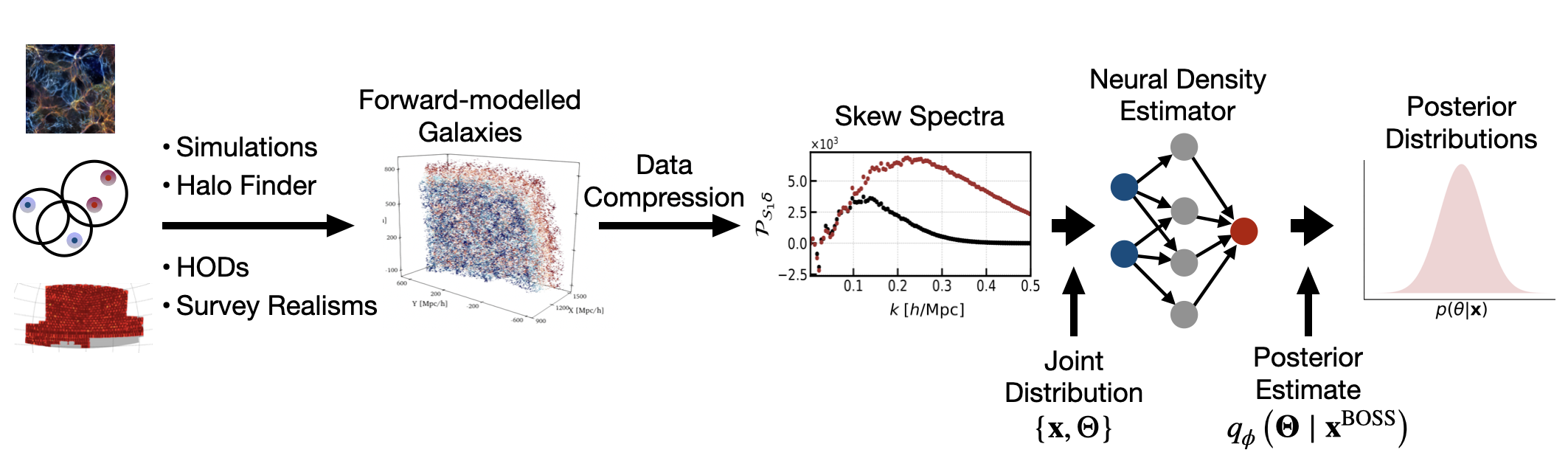}
    \caption{Schematic view of the \simbig \ pipeline; to generate synthetic data that is statistically indistinguishable from BOSS observations, we utilize 2,000 $N$-body simulations from the \textsc{Quijote} suite run at different cosmologies, apply Rockstar halo finder to identify halos, populate halos with galaxies using an extended (9-parameter) HOD model, and finally apply the survey realism. The resulting 20,000 synthetic mocks are then used to train normalizing flows as neural density estimators to obtain approximate posterior distributions, $q_{\phi}$, for cosmological and HOD parameters from given summary statistics ${\bf x}$, which in this paper refer to the skew spectra).}
    \label{fig:pipeline}\vspace{0.1in}
\end{figure*}

\section{\simbig \ Pipeline}
\label{sec:simbig}

\simbig is a framework for performing SBI to infer cosmological constraints from galaxy clustering data, leveraging high-fidelity simulations to forward model the observations and deep generative models to perform inference by estimating the posterior of model parameters. Fig.~\ref{fig:pipeline} shows a schematic sketch of the pipeline. In this section, we describe in more detail the different components of the pipeline, including the forward model, the SBI, and the validation tests.

\subsection{Forward Model}

The current implementation of the \simbig forward model is specifically tailored to produce galaxy catalogs that match the statistical properties of the BOSS-CMASS sample. This involves four integral components: nonlinear dark matter distributions from $N$-body simulations, a halo finder to construct halo catalogs, an HOD model to populate halos with galaxies, and survey realism (including fiber collisions and survey geometry). 

\simbig uses $N$-body simulations from \textsc{Quijote} (high-resolution) suite, featuring 2,000 $\Lambda$CDM cosmologies run in a Latin-hypercube configuration for parameters $\{\Omega_{\rm m}, \Omega_{\rm b}, h, n_{\rm s}, \sigma_8\}$ \citep{Villaescusa-Navarro:2019bje}. The simulations are run with Gadget-III TreePM+SPH code~\citep{Springel2005gadget}, evolving $1024^3$ dark matter particles in a periodic box of $L_{\rm box}=1 \, \gpch$ on the side. The particles are evolved from redshift $z=127$ to $z=0$, with the initial condition generated from the second-order Lagrangian Perturbation Theory (2LPT). Then, we construct halo catalogs using the {\sc Rockstar} halo finder~\citep{Behroozi2012:rockstar}, which identifies halos using phase space information of the dark matter particles.

To populate halos with galaxies, \simbig \ employs the HOD framework \citep{Berlind:2001xk}. We consider a state-of-the-art HOD model \citep{Zheng:2007zg} that includes the standard parameters for characteristic mass scale for halos to host a central galaxy, $\log M_{\rm min}$, the scatter of halo mass at fixed galaxy luminosity, $\sigma_{{\rm log} M}$, the minimum halo mass for halos to host a satellite galaxy, $\log M_0$, the characteristic mass scale for halos to host a satellite galaxy, $\log M_1$, and power-law index for the mass dependence of satellite occupation $\alpha$. In addition, it also incorporates assembly bias, $A_{\rm bias}$, concentration bias of satellite, $\eta_{\rm conc}$, velocity biases for central, $\eta_{\rm cen}$, and satellite, $\eta_{\rm sat}$, galaxies. For each halo catalog, we generate 10 galaxy catalog with different values of the HOD parameters. In total, our training dataset consists of 20,000 galaxy mock catalogs.

The final step is to apply the survey realism. This includes imposing the angular footprint of the survey and the veto masks (including masking for bright stars, centerpost, bad fields, and collision priority). We also model the fiber collision effect for galaxy pairs within an angular resolution $<62''$.\footnote{The overlapping tiling is approximated by randomly downsampling the collided galaxies by 40\%.} In the radial direction, we apply the redshift cut $0.45 < z < 0.6$. 

In total, the forward model involves 5 cosmological parameters and 9 HOD parameters. We chose uniform prior on all parameters except for the assembly bias, where a Gaussian prior centered at zero was implemented. The exact  prior range can be found in Table 1 in~\cite{Hahn2022:simbigMockChallenge}. 
The wide range of the HOD parameter priors results in a large variation in the galaxy number density and clustering amplitude. 

While the \simbig\ pipeline offers a random catalog that is 40 times the size of the real BOSS data sample, to avoid noise induced by the random catalog, we generated an additional random catalog that is 35 times the size of the training set with the highest total galaxy counts. We use the {\sc make survey} toolkit\footnote{\url{https://github.com/mockFactory/make_survey}} to trim the angular footprint of the random.

In order to improve the training and reduce the dynamic range of the data vector, we preprocess the skew spectra of the training catalogs, which exhibit substantial variation in mean number density and clustering amplitude. We remove outliers and normalize the skew spectra in order to ensure that all features contribute meaningfully (Appendix~\ref{app:further_studies}). 

\subsection{Simulation-based Inference}
\label{sec:sbi}

The 20,000 forward modeled galaxy mocks are formally samples drawn from the joint probability distribution $p({\bf \Theta}, \bx)$ of the model parameters, ${\bf \Theta}$, and a summary statistic, $\bx$, measured from observational data. Using this training dataset, we estimate the full posterior distribution over the parameters conditional on the observations, $p({\bf \Theta}|\bx)$. In {\sc SimBIG}, the SBI is performed using neural density estimation with normalizing flows (NFs) \citep{cms/1266935020,Tabak2013AFO,Jimenez-Rezende2015:NF}, which enable efficient estimation of the posteriors with a limited number of simulations.

NFs are powerful tools for constructing expressive probability distributions describing the data using a simple base probability distribution, $\pi({\bf z})$, and a chain of trainable smooth bijective transformations (diffeomorphisms), $f$,  to map the base distribution to the target one. The base distribution is often chosen to be a multivariate Gaussian, which is easy to sample from. The transformations $f$ are chosen to have a tractable Jacobian so that the target distribution can be computed from the base distribution via a change of variables. A neural network is trained to obtain the flow (find the transformation $f$) that best approximate the posterior, $q_\phi({\bf \Theta},\bx) \approx p({\bf \Theta}, \bx)$, by minimizing the Kullback–Leibler (KL) divergence between $p({\bf \Theta},\bx) = p({\bf \Theta}|\bx) p(\bx)$ and $q_\phi({\bf \Theta}|\bx) p(\bx)$. In practice, this is achieved by maximizing the log-likelihood over the training set.

For the analysis of the skew spectra,\footnote{In \simbig analyses of the power spectrum, bispectrum and wavelet scattering transforms MAF was used, while in the field-level analysis using CNNs, Neural Spline Flows~\citep[NSF;][]{Durkan2019:nsf} were utilized. For skew spectra, we found that using the NSF has a significantly slower convergence to stable results.} we use a Masked Autoregressive Flow~\citep[MAF;][]{Papamakarios2017:maf} architecture implemented in the \texttt{sbi} Python package~\citep{tejero-cantero2020sbi}.\footnote{\url{https://github.com/mackelab/sbi}} MAF uses Masked Autoencoder for Distribution Estimation ~\citep[MADE;][]{germain2015made} as a building block and by stacking several of them combines normalizing flows with an autoregressive design, which is well-suited for estimating conditional probabilities such as the posteriors. We split the data into training and validation sets with a 90/10 split, and use only the 90\% for training, leaving the rest for validation test presented in \S \ref{sec:res_valid_rank}. For the split, we randomly select from 20,000 simulations, irrespective of their cosmologies and HOD parameters. 

We perform the optimization using {\sc ADAM} \citep{kingma2017adam} and tune the hyperparameters (number of blocks and hidden layers, dropout probability, 
learning rate, batch size, and number of transformations for the normalizing flows) via experimentation using the {\sc Optuna} hyperparameter optimization framework~\cite{akiba2019optuna} to obtain the best validation score.
The details of the configurations for the {\sc Optuna} search are provided in table \ref{tab:param_config}. We train 2,771 models for our baseline configuration with $R=5 \ \mpch$ and $k_{\rm max}=0.5 \ \hmpc$. 
For each additional configurations that we use to study the dependence of our results on various analysis choices, we train 
2,000--3,000 models. 
To ensure stability and robustness of our results against model variations \citep{lakshminarayanan2017simple,Alsing:2019xrx}, we ensemble average (with equal weights) the top 10 models in each configuration, $q_\phi({\bf \Theta}|\bx) = \sum_{i=1}^{30} q_\phi^i({\bf \Theta}|\bx)/10$. Overall, we do not observe a large variability in the top selected models and they provide consistent results. 

\begin{table*}[htbp!]
\centering
\caption{Hyperparameter configuration used for \textsc{optuna} optimization.}
\begin{tblr}{
  cells = {c},
  cell{1}{2} = {b},
  cell{1}{3} = {b},
  vline{2} = {-}{},
  hline{1,8} = {-}{0.08em},
  hline{2} = {-}{},
}
Hyperparameter & Min & Max & Type & Distribution & Step\\
number of transforms & 5 & 11 & int & uniform & N/A\\
number of hidden units & 256 & 1024 & int & log uniform & N/A\\
number of blocks & 2 & 4 & int & uniform & N/A\\
dropout probability & 0.1 & 0.3 & float & uniform & 0.1\\
batch size & 20 & 100 & int & uniform & 5\\
learning rate & $5 \times 10^{-6}$ & $5 \times 10^{-5}$ & float & log uniform & $10^{-6}$
\end{tblr}\vspace{0.15in}
\label{tab:param_config}
\end{table*}

\subsection{\simbig \ Mock Challenge}
\label{subsec:valid_sims}

As in the previous \simbig analyses, we perform a mock challenge 
using additional 2,000 simulated galaxy catalogs to validate the posterior estimates, $q_\phi$. The test simulations are organized in three datasets as introduced in~\cite{Hahn2022:simbigMockChallenge}. The first two test sets ($\test{0}$, $\test{1}$) are constructed from the {\sc Quijote} $N$-body suite, while the third set ($\test{2}$) uses {\sc AbacusSummit} $N$-body simulations~\citep{Maksimova2021:abacus,Garrison2021:abacus}. Below we list a few important differences between the three test simulations;

\begin{itemize}
    \item $\test{0}$: 500 galaxy mocks generated with the same \simbig \ forward model as the training data, but using a set of 100 {\sc Quijote} simulations at a fixed fiducial cosmology,\footnote{{\sc Quijote} fiducial cosmology is set to $\{\Omega_{\rm m}, \Omega_{\rm b}, h, n_{\rm s}, \sigma_8\}=\{0.3175,0.049, 0.6711,0.9624,0.834\}$} and the HOD model with parameters sampled from a narrower distribution compared to the prior distribution. There are 5 galaxy catalogs per cosmology with 9-parameter HOD models. 
    \item $\test{1}$: 500 galaxy mocks built from the same {\sc Quijote} simulation as in $\test{0}$ but applying a different halo finder and a simpler HOD model.
    Halos are identified using friend-of-friend~\citep[FoF;][]{Davis1985:fof} halo finder 5 galaxy catalogs are constructed per cosmology with 5-parameter HOD models.\footnote{The velocity bias of the central galaxies is fixed to $\eta_{\rm cent} = 0.2$ in order to construct galaxy samples with reasonable power spectrum quadrupole.}
    \item $\test{2}$: 1,000 mocks built from the {\sc AbacusSummit} \texttt{base} simulations. 
    The simulations evolve $6,912^3$ particles in a periodic box of $L_{\rm box}=2\,\gpch$. We use 25 realizations of \texttt{base} simulations, which have the same cosmology and differ in their phase.\footnote{The \texttt{base} simulations are run at Planck2018 \citep{Planck:2018vyg} cosmology corresponding to \texttt{base\_plikHM\_TTTEEE\_lowl\_lowE\_lensing} data.} Each simulation is divided into 8 subvolumes to obtain effectively 200 independent boxes. Halos are identified using {\sc CompaSO} halo finder \citep{Hadzhiyska2022:compaso}. For each box, 5 galaxy catalogs are constructed using the standard 5-parameter HOD models. 
\end{itemize}


\section{Results}
\label{sec:results}

{In this section, we begin by presenting the measured skew spectra of the CMASS-SGC galaxy sample. Subsequently, we detail the conducted validation tests and finally present the first cosmological constraints from the analysis of the full set of galaxy skew spectra on an observational dataset including nonlinear scales. We compare the cosmological and HOD constraints with the previous \simbig \ power spectrum (monopole and quadruple) and bispectrum (monopole) results, and describe the impact of a number of analysis choices on the inferred constraints. }

\subsection{Measured Skew Spectra on \\ BOSS-CMASS-SGC Sample}

\begin{figure*}[t]
    \centering
    \includegraphics[width=.85\textwidth]{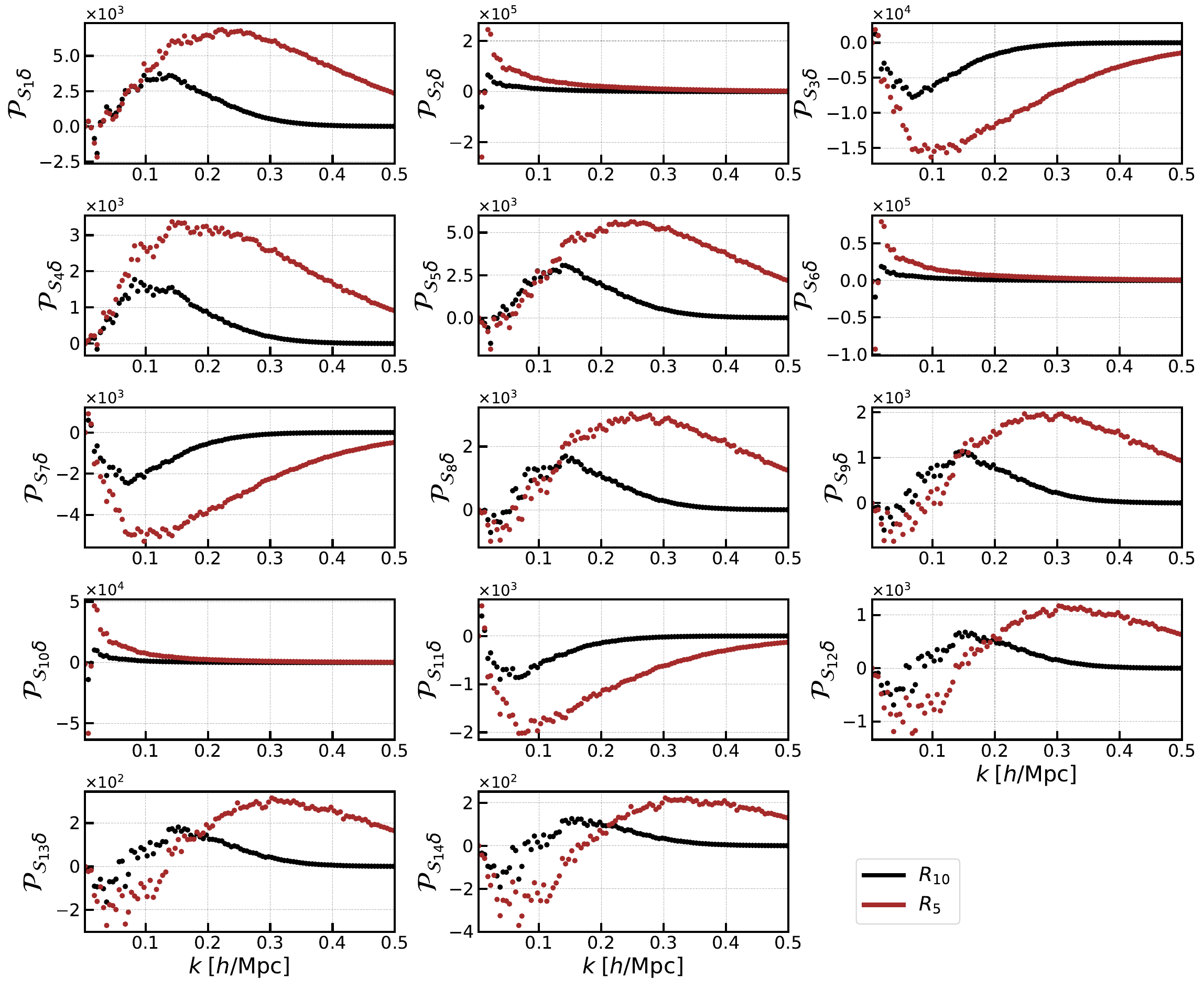}
    \caption{Measurements of the 14 skew spectra from the subsample of the BOSS DR12-SGC galaxies before the shot noise subtraction. Different colors correspond to two smoothing scales of $R=5\,\mpch$ (in red) and $R=10\,\mpch$ (in black). Due to smoothing, skew spectra approach zero at small scales.} \vspace{0.15in}
    \label{fig:skewspec14_obs_R5R10_inclshot}
\end{figure*}

\begin{figure*}[htbp!]
    \centering
    \includegraphics[width=.85\textwidth]{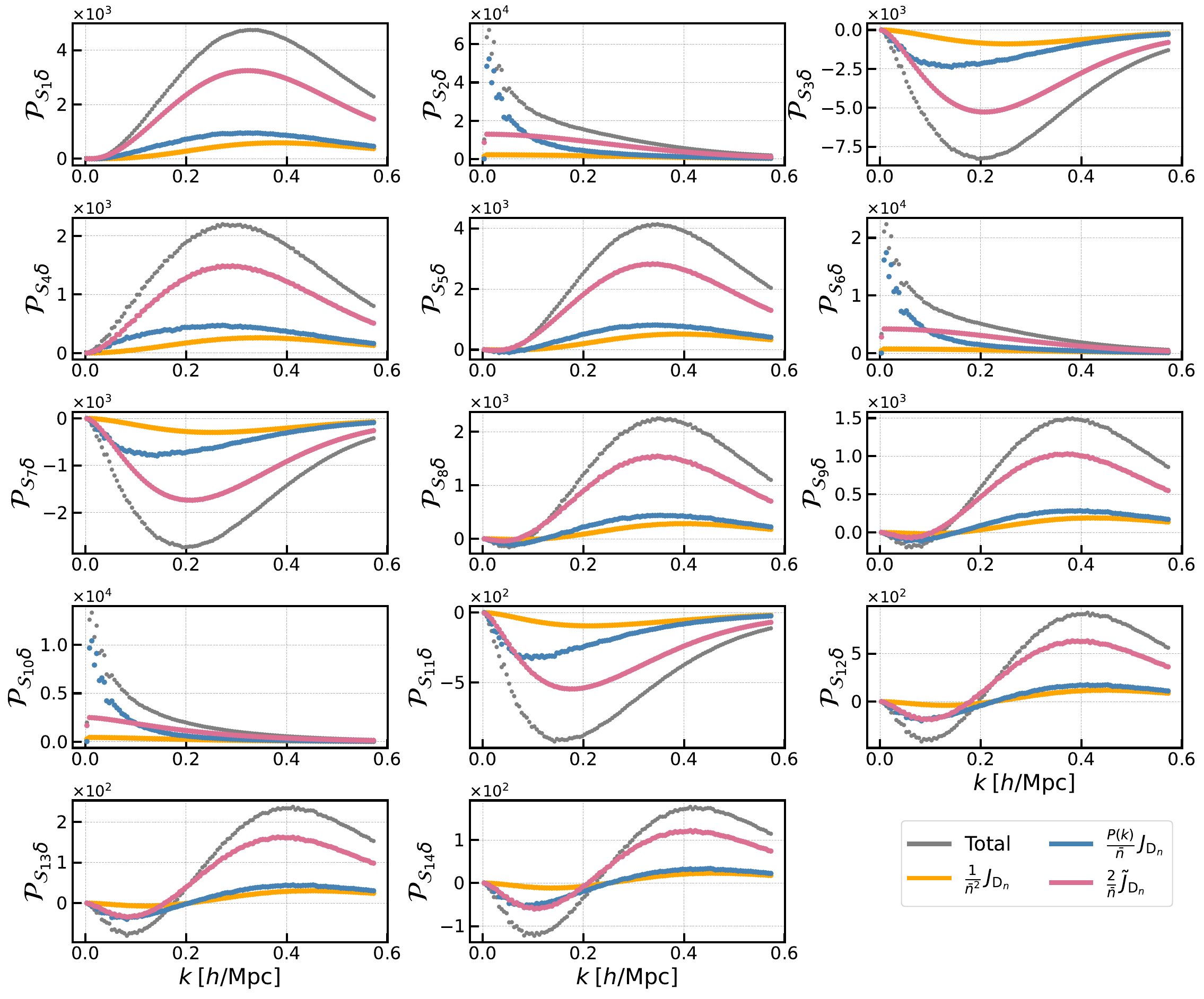}
    \caption{The three contributions to the Poisson shot noise of individual galaxy skew spectra given in Eq.~\eqref{eqn:skewspec_shot}. The smoothing scale is set to $R=5\,\mpch$. Grey curves denote the total shot noise contribution.} \vspace{0.15in}
    \label{fig:shotnoise_skewspec14_R5}
\end{figure*}

Fig.~\ref{fig:skewspec14_obs_R5R10_inclshot} shows the measured skew spectra (including the clustering and shot noise components), applying smoothing scales of $R=5\,\mpch$ in red, and $R=10\,\mpch$ in black. Fig. \ref{fig:shotnoise_skewspec14_R5} shows the Poisson shot-noise contribution, consisting of three contributions in Eq. \eqref{eqn:skewspec_shot}, which have not been previously measured. 

The shapes of the measured spectra are consistent with those on synthetic halo and galaxy catalogs used in \cite{Hou:2022rcd}. As identified before, the skew spectra shapes fall into three categories: (i) ``constant type" involving square of galaxy density field $\delta^2$, and characterized by a large-scale peak and a sharp decline towards smaller scales. (ii) ``displacement type" involving the operator $\partial_i \partial_j\nabla^2$, and distinguished by a positive bump and a zero-crossing feature on large scales for LoS-dependent contributions. (iii) ``tidal type'', involving the tidal operator $S^2$, and marked by a negative dip and anti-correlation with other skew spectra on smaller scales. The skew spectra $\cP_{\cS_{n}\delta}$ with $n>3$ optimally capture the LoS dependence of the bispectrum (see Appendix \ref{app:kernels} for specific forms of the kernels $\cS_n$).

On small scales, the skew spectra all approach zero due to smoothing, as shown in Fig. \ref{fig:skewspec14_obs_R5R10_inclshot}. Comparing Fig.~\ref{fig:skewspec14_obs_R5R10_inclshot} and Fig.~\ref{fig:shotnoise_skewspec14_R5} we also notice that the clustering component is smaller than the Poisson shot noise around $k=0.4\,\mpch$, coincide with the averaged inter-particle separation $\Delta s\sim\,15\,\mpch$ for the BOSS galaxy sample.

\begin{figure*}[t]
    \centering
    \includegraphics[width=.99\textwidth]{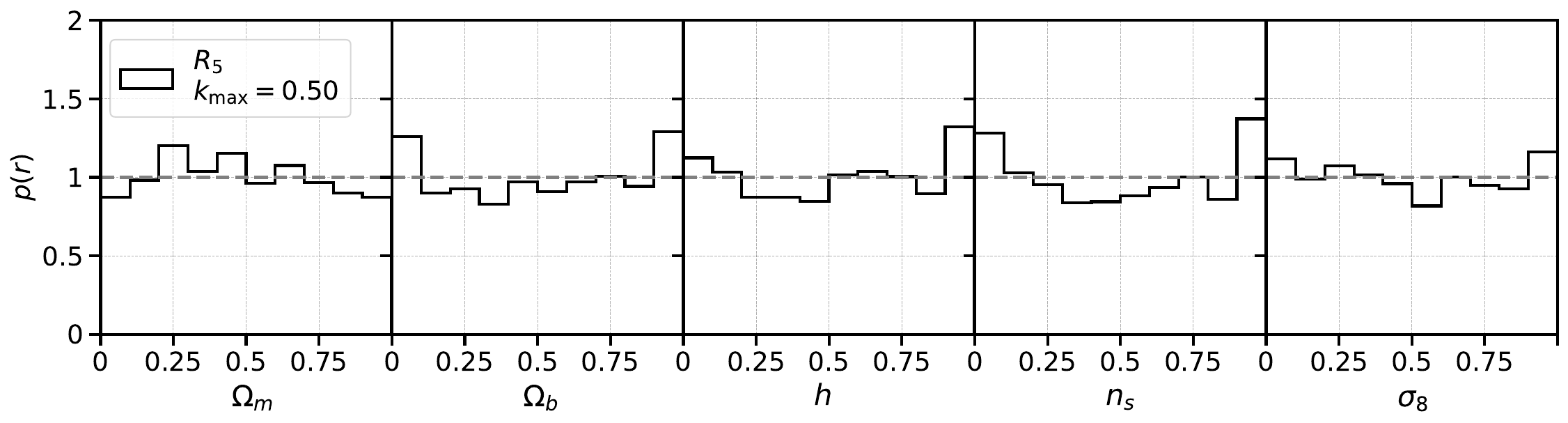}
    \caption{Accuracy test: shown are the rank statistics of the validation dataset for the  configuration $R=5\,\mpch$ and $k_{\rm max}=0.5\,\hmpc$. For most of the parameters, the rank statistics are uniformly distributed. The slight $\cap$-shape in the $\Omega_{\rm m}$ plot is an indication of an under-confident (conservative) error bar.}
    \label{fig:rank_stat_valid_R5_k0x0d5}
\end{figure*}

\begin{figure*}[htbp!]
    \centering
    \includegraphics[width=1.\textwidth]{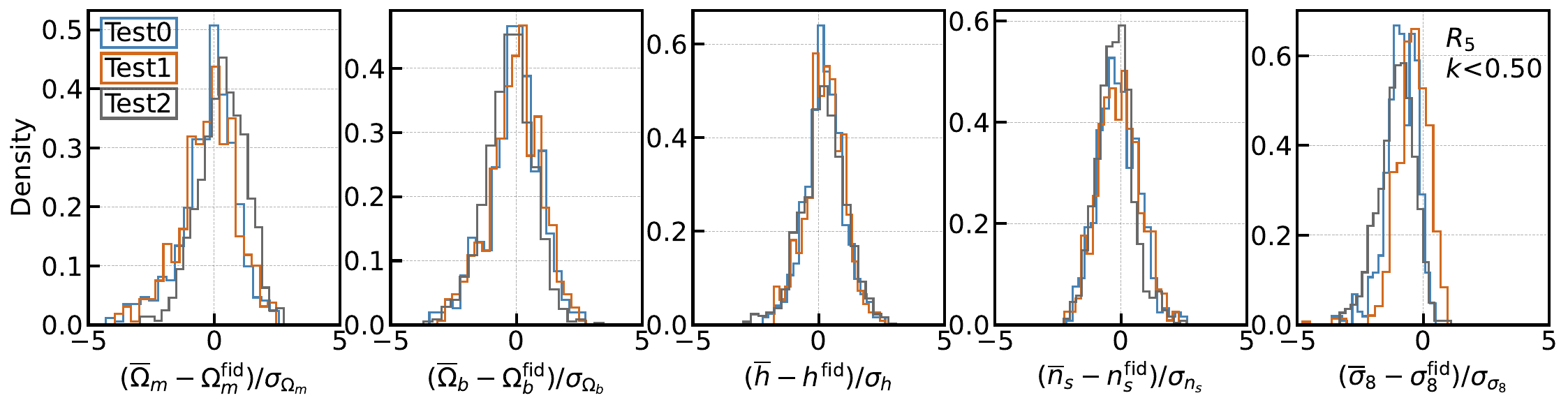}
    \caption{Robustness test: the distribution of the difference between the inferred posterior mean $\bar{\theta}_i$ and the fiducial value ${\theta_i}^{\rm fid}$ normalized by the standard deviation of the posterior $\sigma_{\theta_i}$ for each cosmological parameter on the three sets of test simulations.} 
    \label{fig:valid_posterior_dist_kmax0d5}\vspace{0.15in}
\end{figure*}

\subsection{Posterior Validation}

Before applying the estimated posterior, $q_\phi$, on the skew spectra of the observed CMASS galaxies, we perform two validation tests to asses whether $q_\phi$ can {\it robustly} infer {\it unbiased} posteriors of the cosmological parameters. We describe these tests in this section. 

\subsubsection{Accuracy Test: Simulation-Based Calibration}\label{sec:res_valid_rank}

Our first test aims to assess the accuracy of the approximate posterior, $q_\phi$. Although, theoretically, a sufficiently large number of simulations and optimally trained normalizing flows would inherently guarantee this accuracy by design (as we minimized the KL divergence between $q_\phi$ and the true posterior), \simbig employs a relatively small number of simulations for the size of the data vector 
and dimensionality of the parameter space. Hence, assessing the accuracy of the estimated posterior is pivotal. 
To conduct this evaluation, we employ simulation-based calibration~\citep[SBC;][]{talts2020validating}
on the validation dataset that was excluded from the training of our posterior estimate (see \S \ref{sec:sbi}).

For each test simulation with skew spectra, $\bf{x}_{{\rm test},i}$, we draw $N_{\rm sample}=10,000$ samples from $q_\phi(\boldsymbol{\Theta}\,|\,\bf{x}_{{\rm test},i})$. 
We then, calculate the fraction of the parameters that are below the true cosmological parameter of the simulation: ${\rm rank(\boldsymbol{\Theta}_x)}\equiv  \sum_{i=1}^{N_{\rm sample}}\left[1|_{\boldsymbol{\Theta}_{x,i}<\boldsymbol{\Theta}_x^{\rm fid}}\right]/N_{\rm sample}$. 

Fig.~\ref{fig:rank_stat_valid_R5_k0x0d5} shows the distribution of the rank statistics for $k_{\rm max}=0.5\,\hmpc$ and $R=5\,\mpch$. In the case of independent samples from the true posterior, we should expect uniformly distributed rank statistics. We see that the rank statistics for all parameters are nearly flat. For $\Omega_{\rm m}$, we see a slight $\cap$-shape distributed, which is an indication that the estimated posterior is broader than the true posteriors (i.e. underconfident). Therefore, our constraint should be considered conservative. In Appendix~\ref{app:further_studies} we also show the validation tests for a larger smoothing scale of $R=10 \, \mpch$ and a smaller cutoff of $k_{\rm max} = 0.25 \, \hmpc$. 



\subsubsection{Robustness Test: Mock Challenge}

Next, we perform the \simbig \ ``mock challenge'' using the test simulations $\test{0}$, $\test{1}$, and $\test{2}$ described in \S\ref{subsec:valid_sims}. 
Among the three test simulations, $\test{0}$ is the closest to the training dataset as they share the same $N$-body simulation, halo finder, and HOD model parameters. In contrast, $\test{1}$ used a different halo finder (FoF, which does not resolve substructure as well as the {\sc Rockstar}). It includes only standard HOD parameters and assumes that only the halo mass governs the galaxy–halo connection. Lastly, $\test{2}$ is the most different one from the training dataset; It is built from a different $N$-body simulation ({\sc Abacus}) with halos identified with a different halo finder ({\sc Compaso}) and populated with galaxies using a standard 5-parameter HOD model.

To test the robustness of our trained model to changes in the forward model, we compute the mean and the standard deviation of the parameter posteriors for each test simulation and plot the distribution of the difference between the posterior mean ($\bar {\theta_i}$) and the fiducial value, normalized by the standard deviation $\sigma_{\theta_i}$ for each of the parameters and show the three test sets in Fig.~\ref{fig:valid_posterior_dist_kmax0d5}. If the analysis is robust to variation of the forward model, we expect to obtain a statistical consistency for the distributions across the three test sets, while the posterior mean is not expected to be an unbiased estimate of the true parameter. Given the good consistency for the five cosmological parameters for the three test sets, we conclude that our pipeline is robust and can be applied to observational data.

\subsection{Inferred Posterior on BOSS CMASS-SGC Data}
\label{sec:boss_posterior}

Having validated the trained models on various test simulations, we now move on to applying $q_\phi$ to observational data. {To illustrate the importance of extracting non-Gaussian information of the LSS beyond what is encoded on the power spectrum, we first compare the cosmological constraints from the skew spectra with those from previous \simbig analysis of the power spectrum multipoles \citep{Hahn2022:simbigPk}. Next, we compare the results from the skew spectra with those from previous \simbig analysis of the bispectrum monopole \citep{Hahn2023:simbigBk} to compare their information content. Lastly, we discuss the dependence of the inferred constraints on several of the analysis choices including the smoothing scale, the minimum and maximum scale cuts, the shot noise subtraction. Table \ref{tab:param_config} of Appendix \ref{app:full} presents the 1$\sigma$ uncertainties on cosmological parameters for each of these studies.}

\subsubsection{Comparison with Power Spectrum Multipoles} \label{subsub:Pl}

\begin{figure}[t]
    \centering    
    \includegraphics[width=.48\textwidth]{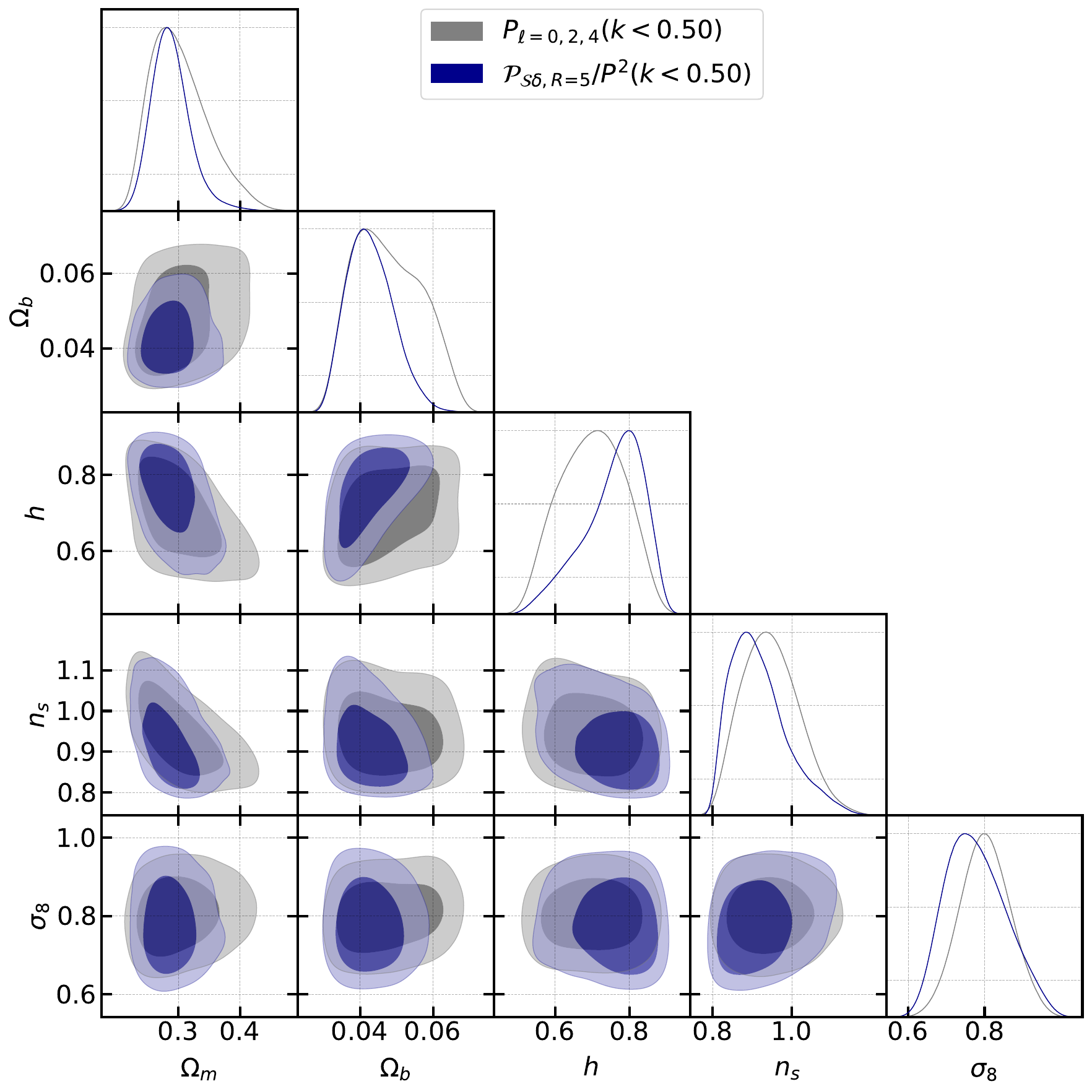}             
    \caption{Posterior distribution of cosmological parameters from the power spectrum multipoles (grey) and the skew spectra (blue) setting $k_{\rm max}=0.5\,\mpch$ for both. The smoothing scale for the skew spectra is set to $R=5\,\mpch$.} 
    \label{fig:contour2d_R5_skewXpk_param5}
\end{figure}

For this comparison, we select an optimal scale in both the maximum wavenumber $k_{\text{max}} = 0.5\,\hmpc$ and a smoothing scale of $R = 5\,\mpch$. This scale choice should, in principle, allow us to extract information from the galaxy field maximally. {We notice that, the $k_{\rm max}$ cut for the skew spectra and the power spectrum (or the bispectrum) is not strictly speaking equivalent. As we will allude below, on the one hand, smoothing of the observed field washes out small-scale information and on the other hand, the convolution of two density fields, mixes small- and large-scale information.}

Fig.~\ref{fig:contour2d_R5_skewXpk_param5} shows the inferred posteriors on cosmological parameters from the skew spectra (blue) and the power spectrum multipoles (grey). The full posterior distributions including the HOD parameters are shown in Fig.~\ref{fig:contour2d_k0d00x0d5_param14_R5_bbn} of Appendix \ref{app:full}. Without applying the BBN prior, we find the posterior mean and the 68\% CF to be $\Omega_{\rm m}=0.288^{+0.024}_{-0.034}$, $\Omega_{\rm b}= 0.043^{+0.005}_{-0.007}$, $h=0.759^{+0.104}_{-0.050}$, $n_{\rm s} = 0.918^{+0.041}_{-0.090}$, and $\sigma_8= 0.778^{+0.066}_{-0.093}$. Compared to the power spectrum, the skew spectra improve the constraints in terms of the 68\% CL by $34\%$, $35\%$, and 18\% in matter density $\Omega_{\rm m}$, baryon density $\Omega_{\rm b}$, and $h$, respectively. However, the constraint in $\sigma_8$ are weaker than that from the power spectrum. As mentioned earlier, the larger uncertainty on $\sigma_8$ can be associated with two facts. First, smoothing the observed field in the skew spectra washes out some information on small scales, which plays a role in constraints on $\sigma_8$. Second, the convolutional nature of skew spectra kernels causes a redistribution of information across various scales. Consequently, the influence of $\sigma_8$ is not exclusively related to small scales. Imposing the BBN prior on baryon density using importance sampling with $\omega_b=\Omega_b / h^2=0.02268 \pm 0.00038$, the skew spectra improve the constraints on matter density $\Omega_{\rm m}$, baryon density $\Omega_{\rm b}$, and $h$, by $34\%$, $49\%$, and 38\%, respectively.

{Despite the reported improvements over the \simbig power spectrum, it is crucial to recognize two points; first, the inferred posteriors from the power spectrum on most parameters are underconfident, thus, the reported relative improvement by the skew spectra can potentially change with a better-trained model for the \simbig power spectrum. Second, the skew spectra utilized in this study are not fully optimized for constraining cosmological parameters. Instead, they represent maximum-likelihood estimators of amplitude-like parameters. Consequently, the reported constraints could potentially be enhanced by strategically weighting the quadratic field using either the Fisher information matrix \citep{Heavens:1999am} or the score function \citep{Alsing:2017var}. The anticipated improvement arises from up-weighting the Fourier modes that contribute the most to the constraints on a given cosmological parameter. We defer further investigation into this possibility to future work.}

\subsubsection{Comparison with Bispectrum Monopole} \label{subsub:B0}

{Given that the construction of skew spectra inevitably mixes scales and washes out the information below the smoothing scale, we compare the skew spectra and the bispectrum using a more conservative choice of scales: applying a smoothing scale of $R=10\,\mpch$ and a lower cutoff scale of $k_{\text{max}} = 0.25\,\hmpc$. The comparison of the constraints from the three statistics at higher $k_{\rm max}$ is presented in Appendix \ref{app:further_studies}.}

Fig.~\ref{fig:contour2d_k0d00x0d25x0d50_param5} shows the constraints from the skew spectra (blue), the power spectrum multipoles (grey) \citep{Hahn2022:simbigPk}, and the bispectrum monopole (magenta)~\citep{Hahn2023:simbigBk}. Here the BBN prior is imposed for all analyses. The constraints from the bispectrum monopole are obtained using $k_{\text{max}} = 0.3\,\hmpc$, while those from the power spectrum assume $k_{\text{max}} = 0.25\,\hmpc$. The three summary statistics provide consistent cosmological constraints. The width of the posteriors from $\cP_{\cS_n}$ and $B_0$, except for $\sigma_8$, are nearly identical. Again, the reduced constraining ability of the skew spectra on $\sigma_8$ can be attributed to its inherent smoothing and convolutional structure, as previously discussed. Without BBN prior, the constraints from the bispectrum monopole on most of the parameters are tighter. We note that since for $R=10 \ \mpch$ the inferred posteriors of all parameters from skew spectra show under-confidence (see Fig. \ref{fig:rank_stat_valid_R10_k0x0d25}), the reported comparison with the bispectrum monopole may be altered with future improvements of \simbig framework.

Although the above configuration offers a more  meaningful comparison between the skew spectra and the bispectrum, we also compare the results of the two statistics for a smaller value of the smoothing scale of $R=5\,\mpch$ and considering $k_{\rm max}=\{0.25,0.5\}\,\hmpc$. With the reduced smoothing scale, the constraints by the skew spectra is weaker when contrasted with the bispectrum. Additionally, we observed a $\sim$1$\sigma$ discrepancy in $\Omega_{\rm b}$ between the skew spectra and the bispectrum monopole under these scale cuts. A similar discrepancy in $h$ emerges when applying the BBN prior. The fact that skew spectra and bispectrum results align only with conservative scale cuts suggests model misspecification may be affecting each statistic differently. This could be due to the skew spectra's sensitivity to redshift-space distortions, which are not entirely accounted for in the bispectrum monopole. In particular, the HOD model's inaccuracies in characterizing satellite galaxy kinematics might affect the results from the skew spectra. A thorough understanding of this model misspecification 
requires more detailed future research.

\begin{figure}[t]
    \centering
    \includegraphics[width=.48\textwidth]{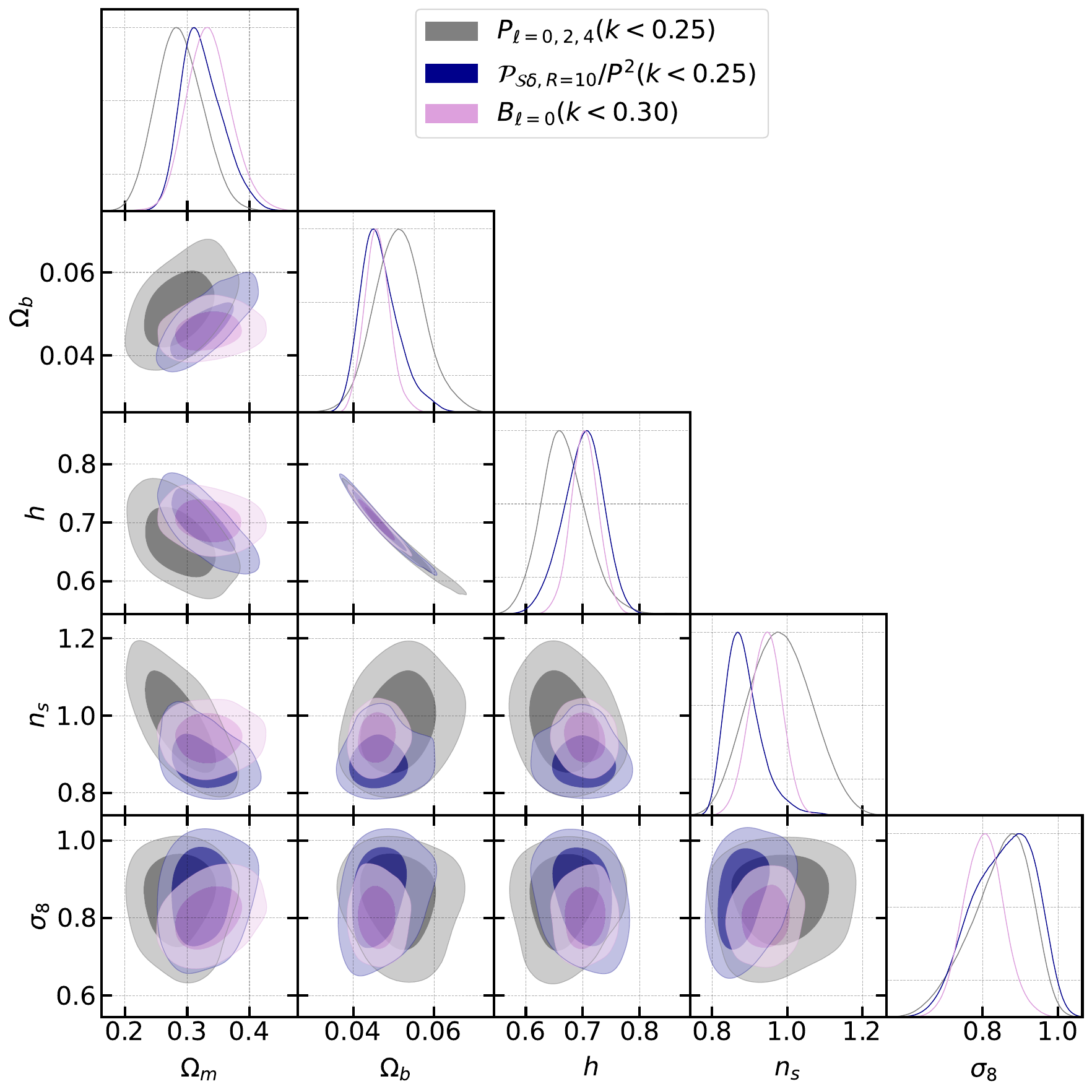}        
    \caption{Posterior distribution of cosmological parameters from the power spectrum (grey), the skew spectra (blue), and the bispectrum monopole (red) imposing the BBN prior, and setting $k_{\rm max}=0.25\,\mpch$ for the first two and $k_{\rm max}=0.3 \ \hmpc$ for the third. The smoothing scale is set to $R=10\,\mpch$ for the skew spectra.}
    \label{fig:contour2d_k0d00x0d25x0d50_param5}
\end{figure}

\subsubsection{Constraints on HOD parameters}\label{subsub:HOD}
In addition to the cosmological parameters, we also show the full posterior distribution including the HOD parameters with BBN priors, in Figs.~\ref{fig:contour2d_k0d00x0d5_param14_R5_bbn} of Appendix \ref{app:full}, respectively. Our analysis reveals no significant enhancement in HOD constraints using the skew spectra compared to the power spectrum multipoles and bispectrum monopole. We found that the posterior distribution of $\sigma_{\log M}$ touches the prior range from the right, indicating that we need to extend the HOD simulations to a wider prior.\footnote{This is the case with or without the BBN prior.} However, this expansion is unlikely to substantially affect the posterior of other parameters, considering the weak degeneracies between $\sigma_{\log M}$ and other parameters. Additionally, the degeneracy directions in HOD parameters derived from the skew spectra are similar to those from the power spectrum multipoles, as predicted by~\cite{Hou:2022rcd}. This similarity is likely attributable to both estimators accessing redshift-space information. 
It is worth noting that among cosmological parameters, $\sigma_8$ is the one most degenerate with HOD parameters. Given the limited constraints on HOD parameters from skew spectra, this degeneracy can be partially responsible for the rather weak constraints on $\sigma_8$. More discussion in the implication of posterior distributions of HODs can be found in Appendix~\ref{app:full}.

\subsubsection{Assessment of Alternative Analysis Configurations} \label{subsub:dep}

To gain further insight into how alternative analysis choices impact the inferred posteriors, we {trained additional sets of normalizing flow models for five additional configurations. For each case, we trained 2,000 -- 3,000 models and considered the ensemble average of the top 10 models with lowest validation loss. All analyses are performed setting the smoothing scale of $R=5 \ \mpch$ and including scales up to $k_{\rm max}=0.5 \, \hmpc$. We summarize our observations here and refer the interested reader to Appendix \ref{app:further_studies} for further details and visualizations of the results. We also present the 1$\sigma$ constraints for these analyses in Table \ref{tab:posterior_summary}}
\begin{itemize}[left=0em]
    \item {Removal of the outliers in the training data: we remove training galaxy catalogs with extreme number densities and skew spectra amplitudes before conducting SBI. Removing these outliers changes the constraining power of the inferred posteriors by $<15\%$. This consistency holds both with and without the BBN prior (see Appendix \ref{app:preprocess} for further details on different outlier removal schemes).}
    \item Subtraction of the Poisson shot noise: {in the standard analysis based on an explicit form of the likelihood, the Poisson shot noise is subtracted in the estimator of a given summary statistics to reduce the covariance. With our SBI approach,} the inferred posteriors with or without shot noise subtraction in the skew spectra estimator yields similar constraints (see Fig.~\ref{fig:contour2d_R5_shot_scales_param5}). This suggests that the SBI method effectively distinguishes the shot noise and clustering components of the skew spectra. Further validation of this interpretation for galaxy samples of varying number densities is left for a future work. 
    \item {Choice of the large-scale cutoff, $k_{\rm min}$:} given that the systematic weights for BOSS data are not well-tested on large scales, we investigate the impact of {excluding the scales close to the survey's fundamental modes $k_{\rm f}$ from the analysis.} We find that setting $k_{\rm min} = 0.01 \ [\hmpc]$, as done in the standard BOSS analyses ({\it e.g.} \cite{BOSS:2016psr}), has only a percent level impact on the overall constraints on $\Lambda$CDM parameters. This is somewhat expected since imposing this cutoff only excludes a single $k$-bin given our binning scheme.
    \item {Choice of the small-scale cutoff, $k_{\rm max}$: to quantify the dependence of the results on the small-scale cutoff,} we experimented with $k_{\rm max} = \{0.15, 0.25, 0.5\} \, [\hmpc]$. The most significant degradation, especially in $\Omega_{\rm m}$ (30\%), was observed with a cutoff $k<0.15\,\hmpc$. Even with such an aggressive cutoff, the {overall} constraints remained comparable to those from the power spectrum {{with $k_{\rm max} = 0.5 \, \hmpc$}. This conclusion holds with or without BBN priors (see Fig. \ref{fig:contour2d_R5_shot_scales_param5})}. 
    \item Choice of smoothing scale: we compared the results with the posterior inferred from skew spectra with smoothing scale $R=10\,\mpch$. 
    Larger smoothing scales leads to expected degradation in constraints, most notably a 40\% decrease in $\Omega_{\rm b}$ precision. Other parameters were less affected {(see Fig. \ref{fig:contour2d_R5R10_param5}).} \\
\end{itemize}

\section{Discussions and Conclusions}
\label{sec:conclude}

This paper presents the first cosmological constraints from analysing the full set of galaxy skew spectra obtained from the BOSS DR12 CMASS-SGC dataset. This work is part of a series of articles by the \simbig collaboration, in which several summary statistics of CMASS galaxy sample were analyzed \citep{Hahn2023:simbigWave2}, including the power spectrum \citep{Hahn2022:simbigMockChallenge, Hahn2022:simbigPk}, the bispectrum \citep{Hahn2023:simbigBk}, the wavelet scattering transform \citep{Blancard2023:simbigWST}, and the field-level analysis using the CNN \citep{Lemos2023:simbigCNN}.  

In order to apply the skew spectra to observational data from spectroscopic galaxy surveys, we developed a new FFT-based estimators for the skew spectra, akin to power spectrum and bispectrum FKP estimators. This estimator integrates essential components such as the survey mask, systematic weights, and a subtraction of the Poisson shot noise.

We use the \simbig framework to perform a simulation-based inference using normalizing flows for neural density estimation to obtain cosmological constraints. The \simbig pipeline incorporates accurate forward modeling of the observed galaxy distribution for the BOSS CMASS-SGC sample, enabling SBI to infer posteriors for cosmological and HOD parameters from summary statistics of choice. Before applying the \simbig framework to CMASS data, as in previous \simbig analyses, we perform two validation tests to ensure that the inferred posteriors are robust and unbiased. Our validation tests, which include simulation-based calibration and the \simbig mock challenge, confirms that the inferred posterior is robust and unbiased for both parameters. 

Setting $k_{\rm max} = 0.5\ \hmpc$ and $R=5 \ \mpch$, and without introducing any informative priors, the skew spectra enhance constraints on $\Omega_{\rm m}$, $\Omega_{\rm b}$, and $h$ by 34\%, 35\%, and 18\%, respectively, over constraints from the power spectrum multipoles. Imposing the BBN prior, we improve the constraint on $h$ by a factor of 2.3, enabling us to constrain $h$ at 3\% level with only 10\% of the full BOSS survey. These relative improvement can potentially be modified given the current underconfident posterior distribution inferred from the \simbig power spectrum.
The inferred value of $n_s$ tends to be lower than that of Planck2018~\citep{Planck:2018vyg}, while the mean of the posterior of $h$ tends to be higher. We attribute these tendencies to either statistical fluctuation or potential model misspecification on small scales. A more detailed investigation of this matter is reserved for future work.


The skew spectra did not yield a reduction in uncertainties on $\sigma_8$ and HOD parameters compared to the power spectrum. This lack of improvement can be attributed to the smoothing and convolutional nature of the skew spectra. The smoothing tends to diminish small-scale modes, and the convolution mixes scales, transferring information from small to larger scales. Consequently, the distinctive information typically associated with $\sigma_8$ and HOD parameters, often tied to small scales, becomes dispersed across various scales, complicating interpretation. To validate this intuition, one could explore an alternative parameterization of the galaxy-dark matter relation, such as using a perturbative forward model \citep{Tucci:2023bag}. Additionally, the skew spectra's limited constraining power may be influenced by the low mean number density and volume of the CMASS-SGC galaxy sample. Further investigation into these issues is deferred to future research. 


{Our results confirm the expectation that on the semi-linear scales (setting $R=10 \, \mpch$ and imposing $k_{\rm max}=0.25 \, \hmpc$), the skew spectra capture the majority of the information of the bispectrum. The \simbig analyses of the bispectrum monopole and the skew spectra yield comparable constraints on all parameters, except for $\sigma_8$, which is better constrained by the bispectrum monopole. Extending the analysis to the smaller scales (setting $R=5 \, \mpch$ and imposing $k_{\rm max}=0.5 \, \hmpc$), we observe the inferred constraints from the skew spectra are weaker relative to the bispectrum. In addition, we report a $\sim1\sigma$ discrepancy in the mean of the posterior of $\Omega_{\rm b}$ and $h$ from the bispectrum monopole and skew spectra. This discrepancy can hint at a potential model misspecification on small scales, possibly due to inaccuracies in HOD descriptions. We leave a more in-depth investigation of this issue to future works.}

Beyond the main result, we performed a series of tests to explore whether our cosmological constraints are impacted by the removal of outliers from training data, the subtraction of the Poisson shot noise, the choice of large- and small-scale cutoff scales, and the choice of the smoothing scale. We found that removal of the outlier can affect the constraints mildly, less than 15\%. Regarding the shot noise subtraction, we observe a marginal impact on the posterior distributions. We found that applying a highly conservative scale cut ($k_{\rm max} = 0.15 \ {\rm Mpc}^{-1}h$) and using a larger smoothing scale lead to  degradation of the constraints on $\Omega_{\rm m}$ and $\Omega_{\rm b}$, respectively, leaving the uncertainties on other parameters largely unaffected. 

This work can be extended in several directions. Firstly, there is room for improvement in the optimality of the skew spectra. The current kernels defining the quadratic fields were designed for optimal constraint on the amplitude of the primordial power spectrum and bispectrum, the growth rate of structure, and galaxy biases. However, they may not fully capture all the shape information. A future extension can refine the kernels to be ``shape-sensitive" and optimally capable of capturing cosmological information.Additionally, the 14 galaxy skew spectra can be broadly classified into three families (see Fig.~\ref{fig:skewspec14_obs_R5R10_inclshot}). Consequently, the size of the data vector may be potentially reduced by constructing optimal combinations.

Given that the skew spectra are optimally designed for constraining PNG, another natural direction for future work is to analyze BOSS data to constrain the Gaussianity of initial conditions using the full set of galaxy skew spectra. Within \simbig framework, this analysis requires incorporating non-Gaussian initial conditions into the forward model and investigating whether the HOD model is affected by PNG.

Combining the power spectrum and the skew spectra should improve the constraints from the two statistics individually. 
Hence, an investigation of an optimal strategy for effectively combining multiple summary statistics is left for future work.

Lastly, considering the observed limited sensitivity of the skew spectra to HOD parameters in our analysis, an interesting avenue for exploration would be to investigate the constraining power of skew spectra when performing SBI using an alternative forward model {\it e.g.,} based on perturbation theory \cite{Tucci:2023bag}. \\

While finalizing this draft, a related work, ~\cite{Chen2024:skewspecBoss}, appeared on arXiv, in which an application of a subset of the skew spectra considered here (neglecting the line-of-sight-dependent quadratic kernels) to BOSS data was presented. In contrast to this work, in which we focused on constraining $\Lambda$CDM using SBI, \cite{Chen2024:skewspecBoss} employed a subset of three skew spectra to constrain primordial non-Gaussianity of equilateral and orthogonal shape using perturbation theory and asssuming a Gaussian likelihood. 

\section*{Acknowledgements}

We would like to thank Fabian Schmidt and Zvonomir Vlah for insightful discussions. JH has received funding from the European Union’s Horizon 2020 research and innovation program under the Marie Sk\l{}odowska-Curie grant agreement No 101025187. AMD acknowledges funding from Tomalla Foundation for Research in Gravity while at University of Geneva, where the majority of this work was carried out. The authors acknowledge the Flatiron Institute Scientific Computing Core and University of Florida Research Computing for providing computational resources and support that have contributed to the research results reported in this publication. \\  \\

\appendix
\vspace{-0.2in}
\section{Explicit Forms of Skew Spectra Kernels in Redshift Space}
\label{app:kernels}

A set of 14 skew spectra, capture the clustering information of the tree-level galaxy bispectrum in redshift space. {Each of the skew spectra corresponds to the maximum likelihood estimator of a specific combination of bias parameters and the growth rate $f$. The explicit forms of the quadratic fields and the parameter combination they are most sensitive to are given below \citep{Schmittfull2020:SSrsd},}
\begin{align}\label{eq:skew_list}
    b_1^3: & \quad \mathcal S_1 = F_2[\delta,\delta],\\
    b_1^2b_2: & \quad \mathcal S_2 = \delta^2, \\
    b_1^2 b_{\mathcal G_2}: & \quad \mathcal S_{3} = S^2[\delta,\delta], \\
    b_1^3f: & \quad \mathcal S_4 = \hat z_i\hat z_j\,\partial_i\left(\delta\frac{\partial_j}{\nabla^2}\delta\right), \\
    b_1^2f: & \quad \mathcal S_5 = 2F_2[\delta^\parallel,\delta] + G_2^\parallel[\delta,\delta], \\
    b_1b_2 f: & \quad  \mathcal S_6 = \delta\delta^\parallel, \\
     b_1 b_{\mathcal G_2} f: & \quad \mathcal S_7 =  S^2[\delta,\delta^{\parallel}], \\
     b_1^2f^2: & \quad \mathcal S_8 = \hat z_i\hat z_j\partial_i\left(\delta\frac{\partial_j}{\nabla^2}\delta^\parallel
    +2\delta^\parallel \frac{\partial_j}{\nabla^2}\delta
    \right),\\
    b_1f^2: & \quad \mathcal S_9 = F_2[\delta^\parallel,\delta^\parallel] + 2 G_2^\parallel[\delta^\parallel,\delta],\\
    b_2f^2: & \quad \mathcal S_{10} = \big(\delta^\parallel\big)^2,\\
    b_{\mathcal G_2} f^2: & \quad  \mathcal S_{11} = S^2(\delta^{\parallel}, \delta^{\parallel}),\\
    b_1f^3: & \quad \mathcal S_{12} = \hat z_i\hat z_j\partial_i\left(\delta^{\parallel\parallel}\frac{\partial_j}{\nabla^2}\delta
    +2\delta^\parallel \frac{\partial_j}{\nabla^2}\delta^\parallel
    \right),\\
    f^3: & \quad \mathcal S_{13} = G_2^\parallel[\delta^\parallel,\delta^\parallel],\\
    f^4: & \quad \mathcal S_{14} = \hat z_i\hat z_j\partial_i\left(\delta^{\parallel\parallel}\frac{\partial_j}{\nabla^2}\delta^\parallel \right). 
\end{align}
{The $F_2$ and $G_2$ functions are the standard perturbation theory kernels for matter density and velocity (see \textit{e.g.} Eq. B.13 in~\cite{Desjacques:2016bnm}), while $S^2$ is the Fourier transform of the Galileon operator, capturing the effect of the tidal field,
\begin{equation}
S^2(\bk_1,\bk_2) \equiv \left(\frac{\bk_1\cdot\bk_2}{k_1 k_2}\right)^2 - 1.
\end{equation}
We defined the redshift-space operators as 
\begin{align}
    \mathcal O^\parallel &= 
    \hat z_i\hat z_j\frac{\partial_i\partial_j}{\nabla^2}\mathcal O, \\
    \mathcal O^{\parallel\parallel} &=
    \hat z_i\hat z_j\hat z_m\hat z_n\frac{\partial_i\partial_j\partial_m\partial_n}{\nabla^4}\mathcal O,
\end{align}
and the operators $\mathcal O [a,b]$ that act on arbitrary fields $a$ and $b$ as 
\begin{equation}
  \label{eq:F2Def}
  {\mathcal O}[a,b](\bk) \equiv \int_{\bf q}  \,\frac{1}{2}\Big[a(\bq)b(\bk-\bq)+b(\bq)a(\bk-\bq)\Big] \, {\mathcal O}(\bq,\bk-\bq). 
\end{equation}
}

\vspace{0.1in}
\section{Pre-processing of the Measurements to remove outliers}\label{app:preprocess}

The \simbig forward-modeled galaxy mocks display wide variations in the mean number density and the clustering amplitude due to the broad HOD parameter priors. To enhance the stability of our training process and ensure that all features meaningfully contribute to the neural networks, we preprocess the training dataset, by removing ``outlier" mock catalogs. Furthermore, we normalize the measured skew spectra by dividing them by square of the power spectrum to obtain features of order unity. We compared several algorithms for removing the outliers, which we describe below.

We remove the outliers based on the amplitude of the skew spectra, and compare the performance of three algorithms; (1) elimination based on the z-score: this is a simple prescription to remove the data points that lie on the tails of a (Gaussian) distribution. The data points with a z-score (the number of standard deviations from the mean) greater than a threshold are declared to be outliers. (2) Local Outlier Factor~\citep[LOF][]{Breunig2000:LocalOutlier} is a machine-learning-based anomaly detection algorithm. It hinges on the idea that anomalies will have a significantly lower density of neighbours than on average, indicating that they are relatively more isolated. It assesses the local density of data points by comparing the density of a data point to its neighbours. (3) Isolation Forest~\citep[IF][]{Liu2009:IsolationForest} is also machine-learning based; the IsoForest isolates anomalies by randomly partitioning the data and creating isolation trees (a decision tree for anomaly isolation). It exploits the fact that anomalies are fewer and more isolated than normal instances, making them easier to separate within the isolation trees.

In the main analyses presented in this paper, we apply the Local Outlier algorithms using its implementation in {\sc sklearn} package~\citep{Pedregosa2011:scikitlearn}. We also tested the impact of not removing the outliers and found that it can change the constraints on the 5 cosmological parameters less than 15\%. The results are summarized in Table~\ref{tab:posterior_summary}. 

\vspace{0.1in}
\section{Full Posterior Distribution: Constraints on the HOD Parameters}\label{app:full}

\begin{figure}[t]
    \centering
    \includegraphics[width=.92\textwidth]{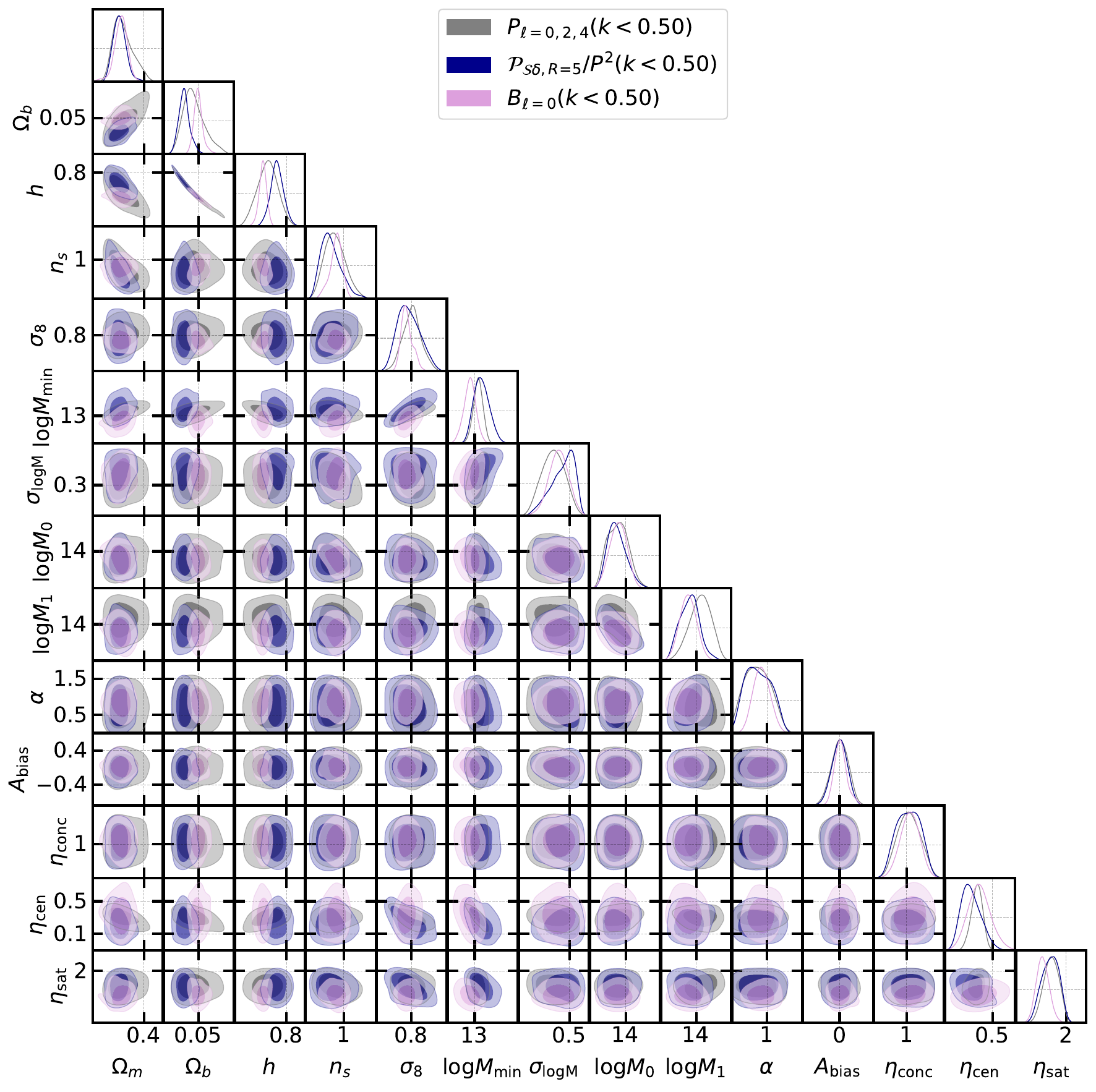}
    \caption{Posterior distribution of cosmological parameters inferred the skew spectra (blue) and power spectrum (grey) with a cut-off scale $k_{\rm max}=0.5\,\mpch$ including the BBN prior and smoothing scale $R=5\, \mpch$.}
    \label{fig:contour2d_k0d00x0d5_param14_R5_bbn}\vspace{0.1in}
\end{figure}

{To illustrate the constraining power of different summary statistics ($P_\ell, B_0, \cS_R$) and the degeneracies between the nuisance and cosmological parameters, in Fig. \ref{fig:contour2d_k0d00x0d5_param14_R5_bbn}, we show the full posterior distributions on model parameters from analyses of the BOSS CMASS-SGC sample. For all three statistics, the BBN prior is imposed, and the small-scale cutoff scale is set to $k_{\rm max} = 0.5 \,\hmpc$. For the skew specra, the smoothing scale of $R=5\, \mpch$ is imposed.}

{Generally, except for $\log M_1$ (the characteristic halo mass scale to host satellite galaxies), the skew spectra do not improve the constraints on any of the HOD parameters over the power spectrum. The inferred posterior on $\sigma_{\rm log M}$ (the scatter in the halo mass - galaxy luminosity relation) hits the prior range, indicating that additional mock catalogs sampling a wider prior range are necessary. Similar to $\sigma_8$, the failure of the skew spectra in improving the constraints on the HOD parameters can potentially be associated smoothing of the input field and the convolutional nature of the skew spectra. The former washes out small-scale modes, while the latter mixes different scales and transfers information from small to larger scales.} This interplay potentially disperses the information usually linked to $\sigma_8$ and the HOD parameters, which are typically associated with small scales, thereby complicating their interpretation. 

The comparison between the skew spectra and the bispectrum monopole reveals that their constraining power on HOD parameters is quite similar. Notably, certain HOD parameters, like $\log M_{\rm min}$ and $\eta_{\rm cent}$, even show enhancements in constraint accuracy. This observation suggests that the velocity information encapsulated by the RSD could be crucial for more effectively constraining HOD parameters.

{Worth noting that among cosmological parameters, $\sigma_8$ shows the strongest degeneracy with several HOD parameters, especially, $\log M_{\rm min}, \eta_{\rm cent}, \eta_{\rm sat}$, which are poorly constraints by the skew spectra. These degeneracies play a part in the limited constraining power of the skew spectra on $\sigma_8$. Therefore, analysis of skew spectra using a different model of dark matter - galaxy relation, {\it e.g.}, the perturbative biasing description, could potentially provide better constraints on $\sigma_8$.}


\vspace{0.2in}
\section{Validation Test with Conservative Scale Cuts}\label{app:SBC_alt}

{In \S \ref{subsub:B0}, we presented a comparison of the skew spectra, the power spectrum multipoles, and the bispectrum monopole for a rather conservative choice of scale cuts ($k_{\rm max}=0.25\,\hmpc$, $R=10\,\mpch$). This section provides the two validation tests to assess the accuracy and robustness of the estimated posteriors from the skew spectra.}

\begin{figure}[t]
    \centering
    \includegraphics[width=.99\textwidth]{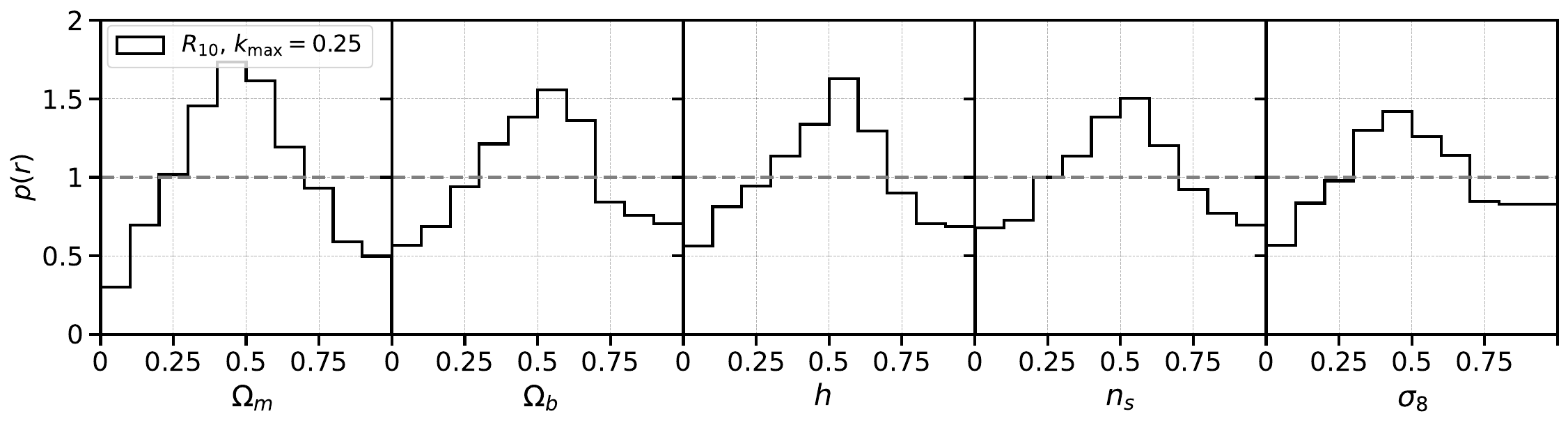}
    \caption{Rank statistics of the validation data for analysis of the skew spectra with $R=10\,\mpch$ and $k_{\rm max}=0.25\,\hmpc$.}
    \label{fig:rank_stat_valid_R10_k0x0d25}\vspace{-0.08in}
\end{figure}

\begin{figure}[htbp!]
    \centering
    \includegraphics[width=1.\textwidth]{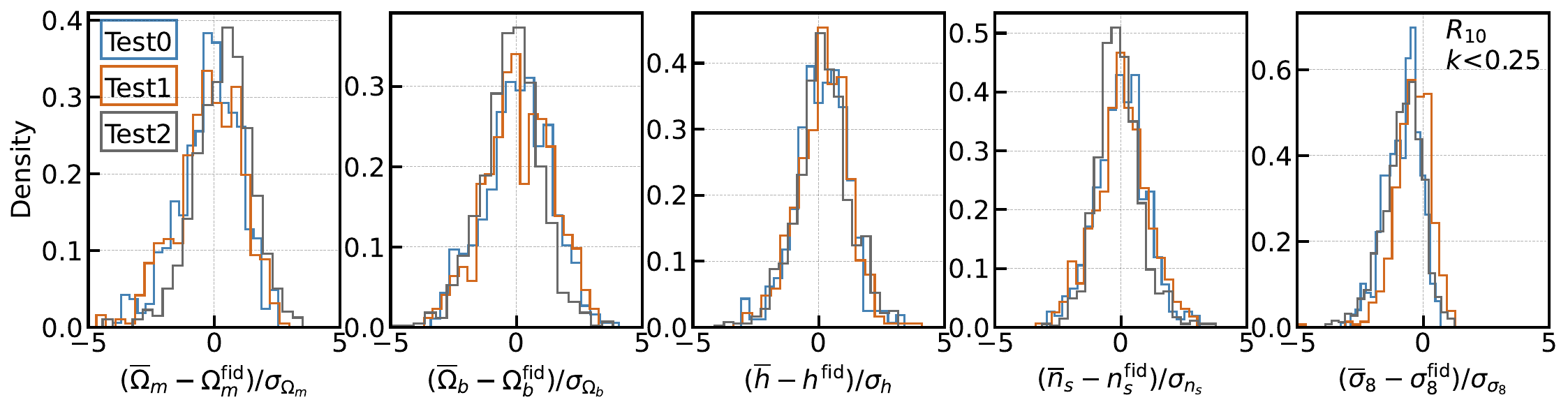}
    \caption{Distribution of the difference between the inferred posterior mean $\bar{\theta_i}$ and the fiducial value ${\theta_i}^{\rm fid}$ normalized by the standard deviation of the posterior $\sigma_{\theta_i}$ for each cosmological parameter on the three sets of test simulations.} 
    \label{fig:valid_posterior_dist_kmax0d25}
\end{figure}

Fig.~\ref{fig:rank_stat_valid_R10_k0x0d25} shows the distribution of the rank statistics, which {exhibits $\cap$-shaped distributions for all parameters.} This behavior indicates that the estimated posteriors are broader than the true posteriors (i.e. underconfident) and implies that the skew spectra constraints should be considered conservative.{Fig.~\ref{fig:valid_posterior_dist_kmax0d25} shows the distributions of the difference between the estimated posterior means and the fiducial values of cosmological parameters across the three test simulations. The statistical consistency of the distributions across the three test sets demonstrates the robustness of the trained models to details of the forward modeling.}

\section{Dependence of the Constraints on Analysis Choices}\label{app:further_studies}


{As described in \S \ref{subsub:dep}, we perform several additional SBI analyses to investigate the dependence on the inferred cosmological constraints on various analysis choices. For each configuration, we re-train 2,000 -- 3,000 neural network models. We report the 1$\sigma$ constraints on cosmological parameters for these tests in Table~\ref{tab:posterior_summary}. We provide further details on these tests here.}

\begin{figure}[htbp!]
    \centering
    \includegraphics[width=.45\textwidth]{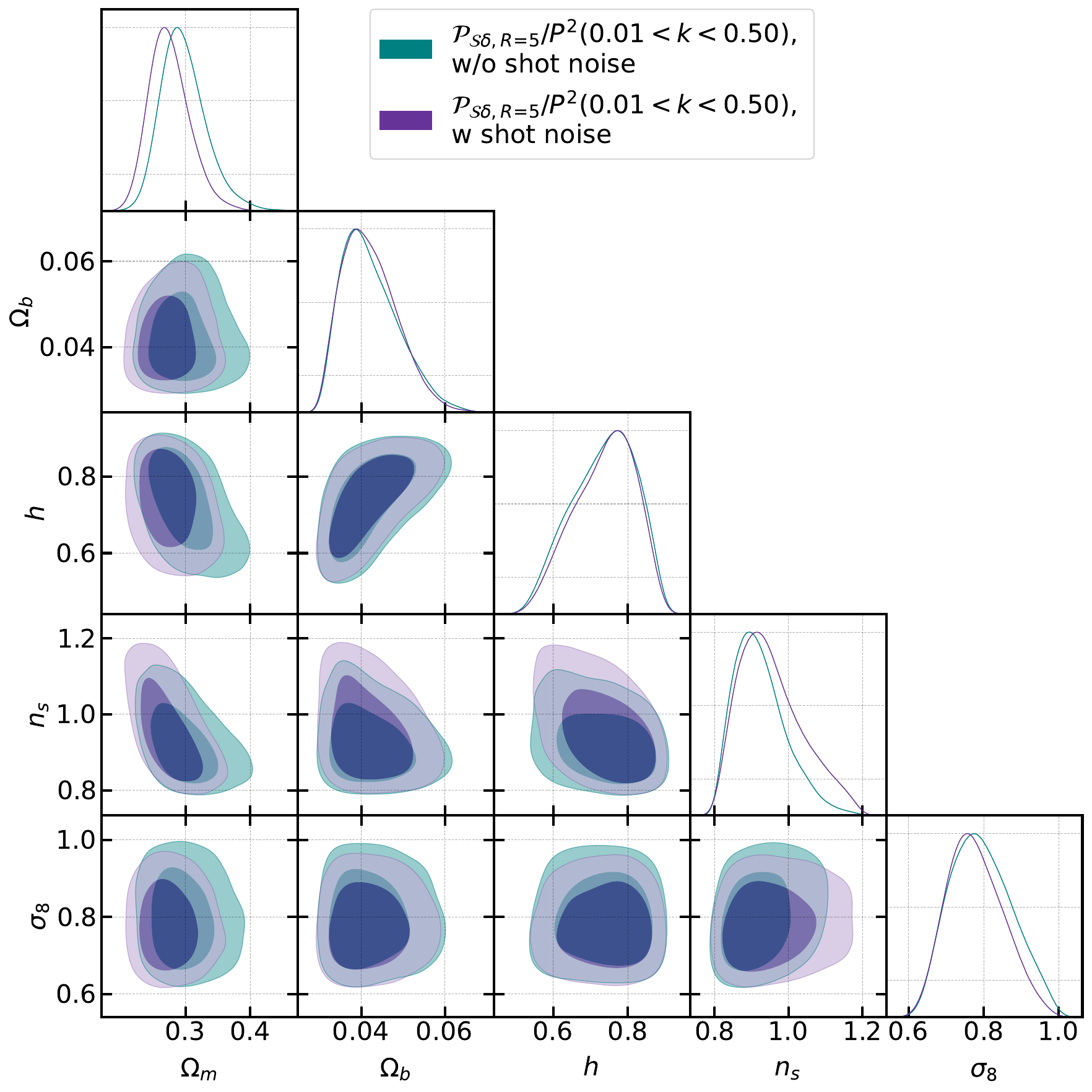}
    \includegraphics[width=.45\textwidth]{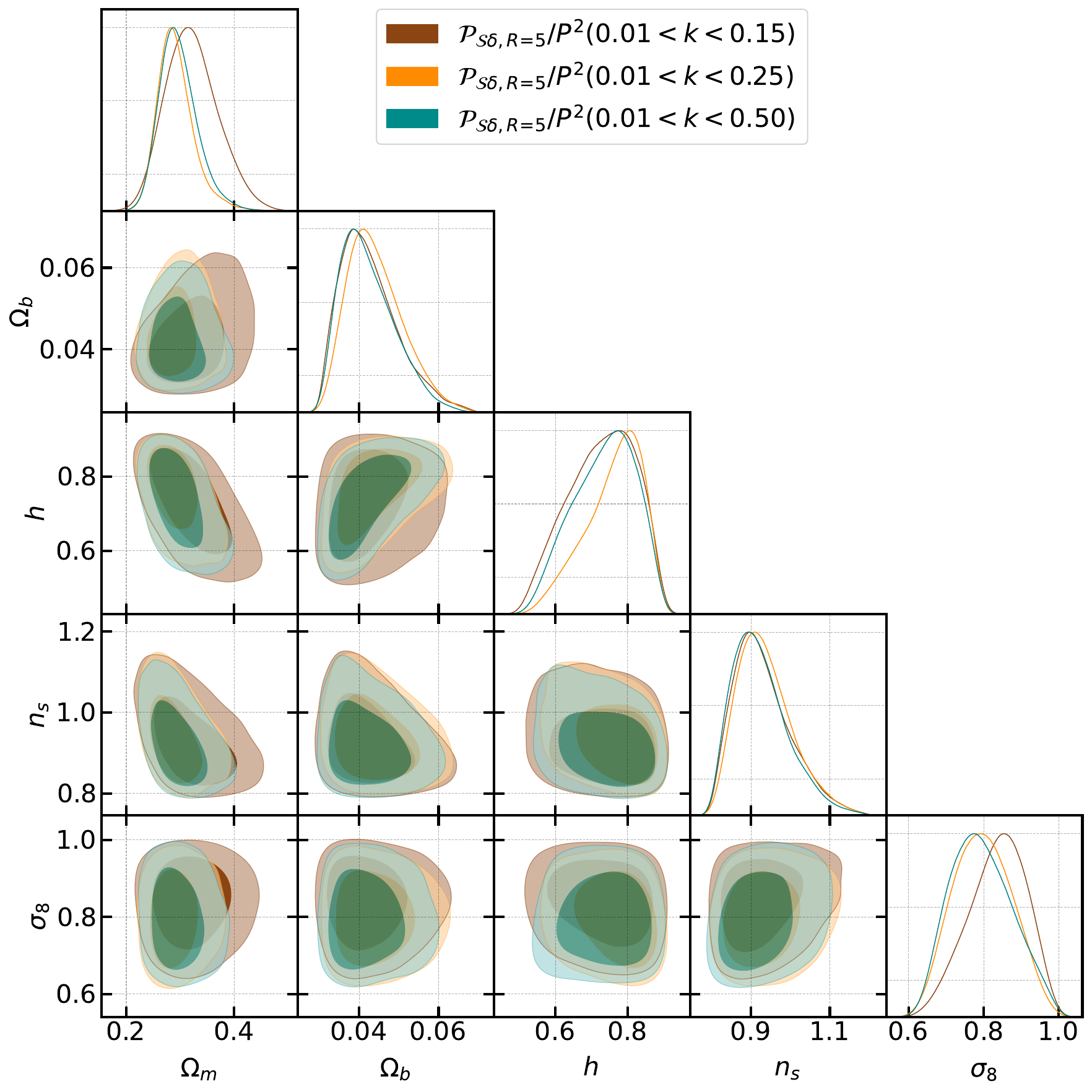}
    \caption{{\it Left:} Posterior distribution of cosmological parameters with (purple) and without (green) subtracting the Poisson shot noise in skew spectra estimators. {\it Right:} Posterior distribution of cosmological parameters for three choices of small-scale cutoff:  $k_{\rm max} = 0.15\,\hmpc$ in red, $k_{\rm max} = 0.25\,\hmpc$ in orange, and $k_{\rm max} = 0.5\,\hmpc$ in green.}
    \label{fig:contour2d_R5_shot_scales_param5}
\end{figure}

First, we study the impact of shot noise. For this test, we opt for $k_{\rm max} = 0.5\,\hmpc$ and $R=5$ to enhance the influence at small scales where shot noise effects are non-negligible. 
The left panel of Fig.~\ref{fig:contour2d_R5_shot_scales_param5} shows the comparison with and without shot noise subtraction. We found that the constraints on $\Omega_{\rm b}$ and $h$ are identical in the two cases, while the widths of the  $\sigma_8$ and $n_s$ posteriors are mildly affected. For $\Omega_{\rm m}$, we observe a slight shift of the peak of the posterior. Despite these marginal differences, we conclude that the constraints are largely unaffected by the subtraction of shot in the estimator. This implies that in contrast to a conventional likelihood-based inference approach, the SBI approach can effectively isolate the influence of shot noise and extract the cosmological response based on a given summary statistics. However, we note that validating this interpretation requires performing further tests on samples with varying number density and employing different summary statistics . 

Second, we investigate the impact of the maximum scale cuts by choosing $k_{\rm max} = \{0.15, 0.25, 0.5\} \, [\hmpc]$. Here, we set the smoothing scale to $R=5\,\mpch$. Due to the convolutional structure of the skew spectra and the smoothing operation, small-scale information is partially transferred to larger scales. As previously shown in the numerical Fisher forecast of ~\cite{Hou:2022rcd}, while skew spectra can achieve better constraints at relatively large scales compared to the power spectrum or the bispectrum monopole, the constraining power also saturates faster as we push the $k_{\rm max}$ cuts towards smaller scales. This saturation is also apparent in the fast-dropping power in  Fig.~\ref{fig:skewspec14_obs_R5R10_inclshot}. As expected, applying an aggressive small-scale cut at $k<0.15\,\hmpc$ degrades the constraints, particularly $\Omega_{\rm m}$ by 30\%. Nevertheless, when imposing the BBN priors, even for this cutoff, the constraints from skew spectra are still comparable to those from the power spectrum multipoles with a scale cut of $k_{\rm max} = 0.5\,\hmpc$. 

\begin{figure}[htbp!]
    \centering
    \includegraphics[width=.44\textwidth]{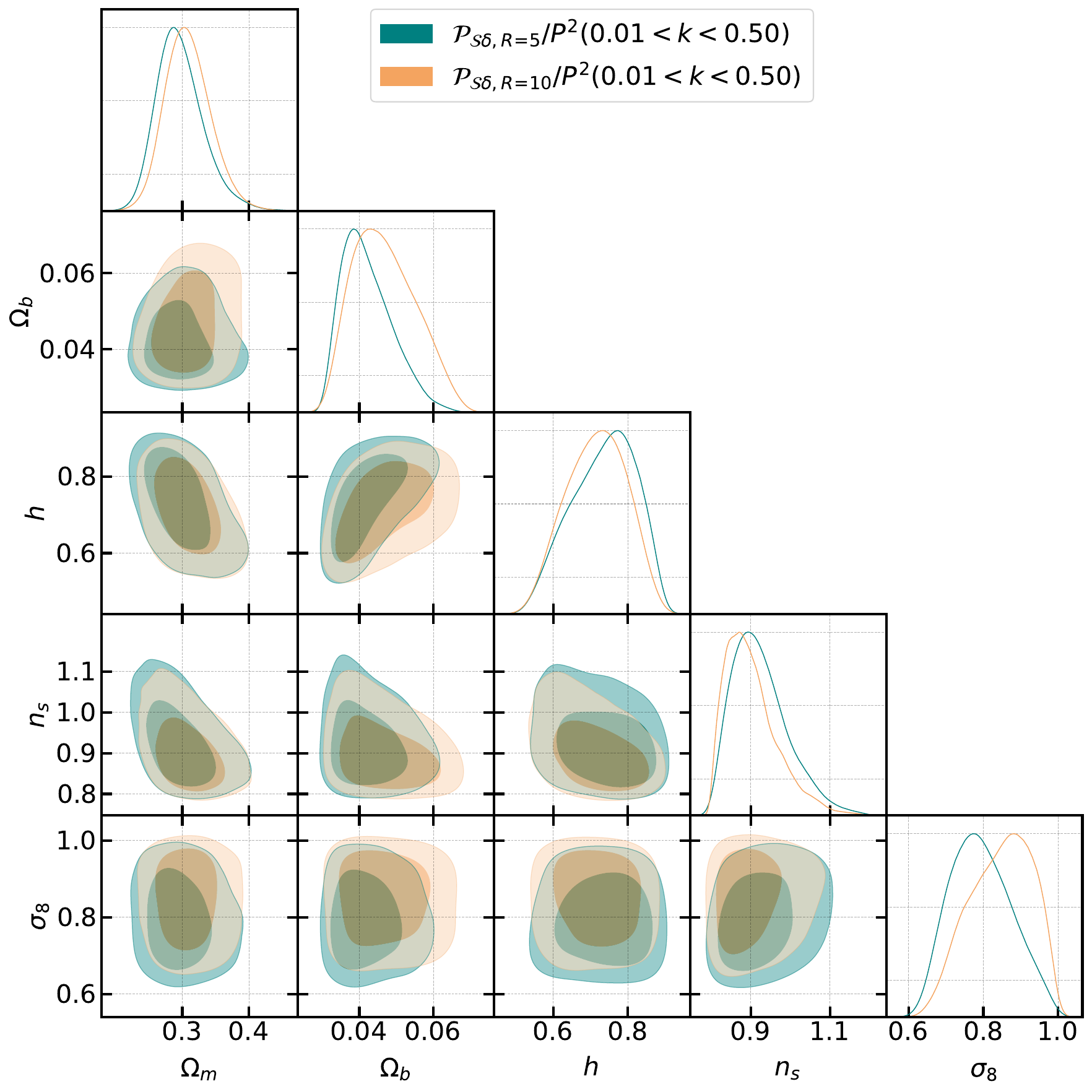}
    \caption{Posterior distribution of cosmological parameters for two choices of the smoothing scale: $R=5 \ \mpch$ in green and $R=10 \ \mpch$ in orange. The small-scale cutoff is set to $k_{\rm max}=0.5 \ \hmpc$ in both cases.}
    \label{fig:contour2d_R5R10_param5}
\end{figure}


Third, we study the impact of the choice of the smoothing scales by imposing $R=5\, \mpch$ and $R=10 \, \mpch$. Fig.~\ref{fig:skewspec14_obs_R5R10_inclshot} shows that the smoothing scale can shift the featured peak of the skew spectra, which could potentially be sensitive to information at different scales. Furthermore, a larger smoothing scale washes out the small-scale fluctuations more significantly. We find that, as expected, the constraints degrade when using a larger smoothing scale. The degradation is particularly visible in $\Omega_{\rm b}$ by $\sim\!40\%$. For other parameters, the effect is marginal. 

\vspace{0.2in}
\begin{table}[htbp!]
\centering
\caption{Table summarizes the posteriors of $\Lambda$CDM cosmological parameters inferred from the skew spectra, and the power spectrum. We present the median and 68\% uncertainties of the parameters.}\vspace{0.1in}
\begin{tblr}{
  column{1} = {c},
  cell{2}{1} = {r=4}{},
  cell{6}{1} = {r=2}{},
  cell{10}{1} = {r=2}{},
  cell{12}{1} = {r=2}{},
  cell{15}{1} = {r=4}{},
  cell{19}{1} = {r=2}{},
  cell{23}{1} = {r=2}{},
  cell{25}{1} = {r=2}{},
  vline{2-3} = {1-26}{},
  vline{3} = {3-4}{},
  hline{2,14} = {-}{0.08em},
  hline{1,27} = {-}{0.09em},
  hline{6, 8-9, 10,12,14-15,19,21-23,25} = {-}{0.05em},
}
no priors & \hspace{0.07in} scale cuts & \hspace{0.2in}   $\Omega_{\rm m}$ & \hspace{0.2in} $\Omega_{\rm b}$ & \hspace{0.2in}   $h$ & \hspace{0.2in}   $n_s$ & \hspace{0.2in}   $\sigma_8$\\
$\cP_{R=5}$ & $k\in (k_{\rm f}, 0.5)$ &  $0.288^{+0.024}_{-0.034}$ & $0.043^{+0.005}_{-0.007}$ & $0.756^{+0.104}_{-0.050}$ & $0.918^{+0.041}_{-0.090}$ & $0.778^{+0.066}_{-0.093}$  & \\
& $k\in (0.01, 0.5)$ &  $0.297^{+0.027}_{-0.040}$ & $0.042^{+0.005}_{-0.009}$ & $0.735^{+0.107}_{-0.072}$ & $0.924^{+0.046}_{-0.087}$ & $0.794^{+0.072}_{-0.098}$  & \\
& $k\in (0.01, 0.25)$ &  $0.292^{+0.024}_{-0.037}$ & $0.044^{+0.005}_{-0.008}$ & $0.761^{+0.102}_{-0.051}$ & $0.936^{+0.050}_{-0.085}$ & $0.798^{+0.078}_{-0.086}$  & \\
& $k\in (0.01, 0.15)$ & $0.322^{+0.042}_{-0.055}$ & $0.043^{+0.005}_{-0.009}$ & $0.729^{+0.118}_{-0.076}$ & $0.930^{+0.047}_{-0.091}$ & $0.837^{+0.091}_{-0.066}$  & \\
$\cP_{R=10}$ & $k\in (0.01, 0.5)$  &  $0.309^{+0.030}_{-0.037}$ & $0.047^{+0.007}_{-0.011}$ & $0.716^{+0.090}_{-0.079}$ & $0.902^{+0.032}_{-0.083}$ & $0.848^{+0.107}_{-0.068}$  & \\
$\cP_{R=10}$ & $k\in (0.01, 0.25)$  &  $0.308^{+0.029}_{-0.036}$ & $0.052^{+0.007}_{-0.007}$ & $0.753^{+0.080}_{-0.063}$ & $0.909^{+0.043}_{-0.079}$ & $0.846^{+0.081}_{-0.079}$  & \\
$\cP_{R=5}$, shot noise incl. &  $k\in (0.01, 0.5)$ &  $0.275^{+0.026}_{-0.036}$ & $0.042^{+0.005}_{-0.008}$ & $0.738^{+0.101}_{-0.068}$ & $0.953^{+0.056}_{-0.111}$ & $0.778^{+0.065}_{-0.087}$  & \\
$\cP_{R=5}$, no outlier removal &  $k\in (0.01, 0.5)$ &  $0.288^{+0.027}_{-0.034}$ & $0.043^{+0.005}_{-0.008}$ & $0.760^{+0.103}_{-0.053}$ & $0.924^{+0.041}_{-0.089}$ & $0.798^{+0.080}_{-0.097}$  & \\
$P_{\ell=0,2}$ \citep{Hahn2022:simbigPk} & $k\in (k_{\rm f}, 0.5)$ & $0.302^{+0.032}_{-0.057}$ & $0.048^{+0.008}_{-0.012}$ & $0.700^{+0.093}_{-0.095}$ & $0.947^{+0.063}_{-0.082}$ & $0.802^{+0.067}_{-0.066}$  & \\
$P_{\ell=0,2}$ \citep{Hahn2022:simbigPk} & $k\in (k_{\rm f}, 0.25)$ & $0.283^{+0.034}_{-0.041}$ & $0.052^{+0.008}_{-0.008}$ & $0.695^{+0.105}_{-0.105}$ & $0.981^{+0.082}_{-0.094}$ & $0.845^{+0.098}_{-0.059}$  & \\
$B_{0}$ \citep{Hahn2023:simbigBk} & $k\in (k_{\rm f}, 0.5)$  &  $0.294^{+0.027}_{-0.026}$ & $0.059^{+0.005}_{-0.005}$ & $0.756^{+0.040}_{-0.040}$ & $0.952^{+0.039}_{-0.034}$ & $0.784^{+0.037}_{-0.042}$ 
& \\
$B_{0}$  & $k\in (k_{\rm f}, 0.3)$  &  $0.336^{+0.034}_{-0.038}$ & $0.045^{+0.004}_{-0.004}$ & $0.696^{+0.041}_{-0.043}$ & $0.944^{+0.043}_{-0.043}$ & $0.802^{+0.051}_{-0.056}$ \\
\hline
BBN prior & \hspace{0.1in} scale cuts & \hspace{0.2in}   $\Omega_{\rm m}$ & \hspace{0.2in} $\Omega_{\rm b}$ & \hspace{0.2in}   $h$ & \hspace{0.2in}   $n_s$ & \hspace{0.2in}   $\sigma_8$\\
$\cP_{R=5}$ & $k\in (k_{\rm f}, 0.5)$ &  $0.285^{+0.028}_{-0.033}$ & $0.041^{+0.003}_{-0.004}$ & $0.750^{+0.034}_{-0.032}$ & $0.915^{+0.047}_{-0.082}$ & $0.781^{+0.071}_{-0.095}$  & \\
& $k\in (0.01, 0.5)$ &  $0.291^{+0.028}_{-0.034}$ & $0.041^{+0.003}_{-0.004}$ & $0.744^{+0.032}_{-0.033}$ & $0.920^{+0.051}_{-0.081}$ & $0.798^{+0.082}_{-0.093}$  & \\
& $k\in (0.01, 0.25)$ &  $0.289^{+0.026}_{-0.034}$ & $0.041^{+0.003}_{-0.004}$ & $0.747^{+0.033}_{-0.035}$ & $0.940^{+0.054}_{-0.083}$ & $0.803^{+0.088}_{-0.085}$  & \\
& $k\in (0.01, 0.15)$ & $0.316^{+0.040}_{-0.053}$ & $0.042^{+0.004}_{-0.007}$ & $0.741^{+0.054}_{-0.044}$ & $0.926^{+0.047}_{-0.079}$ & $0.834^{+0.093}_{-0.072}$  & \\
$\cP_{R=10}$ & $k\in (0.01, 0.5)$  &  $0.306^{+0.030}_{-0.038}$ & $0.045^{+0.004}_{-0.006}$ & $0.712^{+0.044}_{-0.042}$ & $0.898^{+0.040}_{-0.070}$ & $0.853^{+0.098}_{-0.072}$ & \\
$\cP_{R=10}$ & $k\in (0.01, 0.25)$  &  $0.320^{+0.032}_{-0.036}$ & $0.048^{+0.005}_{-0.005}$ & $0.692^{+0.033}_{-0.036}$ & $0.921^{+0.055}_{-0.072}$ & $0.839^{+0.075}_{-0.078}$ & \\
$\cP_{R=5}$, shot noise incl. &  $k\in (0.01, 0.5)$ &  $0.273^{+0.025}_{-0.033}$ & $0.041^{+0.003}_{-0.003}$ & $0.745^{+0.028}_{-0.031}$ & $0.941^{+0.060}_{-0.089}$ & $0.779^{+0.070}_{-0.081}$ & \\
$\cP_{R=5}$, no outlier removal &  $k\in (0.01, 0.5)$ &  $0.285^{+0.026}_{-0.032}$ & $0.040^{+0.003}_{-0.004}$ & $0.751^{+0.033}_{-0.032}$ & $0.924^{+0.050}_{-0.077}$ & $0.802^{+0.098}_{-0.094}$  & \\
$P_{\ell=0,2}$ \citep{Hahn2022:simbigPk} & $k\in (k_{\rm f}, 0.5)$ & $0.301^{+0.034}_{-0.057}$ & $0.047^{+0.005}_{-0.008}$ & $0.701^{+0.055}_{-0.052}$ & $0.941^{+0.062}_{-0.078}$ & $0.804^{+0.062}_{-0.070}$  & \\
$P_{\ell=0,2}$ \citep{Hahn2022:simbigPk} & $k\in (k_{\rm f}, 0.25)$ & $0.286^{+0.037}_{-0.040}$ & $0.051^{+0.007}_{-0.006}$ & $0.669^{+0.032}_{-0.050}$ & $0.985^{+0.086}_{-0.086}$ & $0.849^{+0.092}_{-0.056}$  & \\
$B_{0}$ \citep{Hahn2023:simbigBk} & $k\in (k_{\rm f}, 0.5)$  &  $0.291^{+0.032}_{-0.026}$ & $0.050^{+0.002}_{-0.003}$ & $0.676^{+0.020}_{-0.018}$ & $0.949^{+0.046}_{-0.032}$ & $0.768^{+0.029}_{-0.050}$ & \\
$B_{0}$  & $k\in (k_{\rm f}, 0.3)$  &  $0.335^{+0.032}_{-0.038}$ & $0.046^{+0.003}_{-0.003}$ & $0.704^{+0.024}_{-0.024}$ & $0.943^{+0.043}_{-0.041}$ & $0.803^{+0.051}_{-0.052}$ & \\
\end{tblr}
\label{tab:posterior_summary}
\end{table}

\clearpage
\bibliographystyle{aasjournal}
\bibliography{skewspec_simbig}

\end{document}